\documentclass[amsmath,amssymb,aps,pre,showpacs,twocolumn,groupedaddress]{revtex4-1}
\pdfoutput=1
\usepackage[utf8]{inputenc}
\usepackage{hyperref}
\usepackage{wrapfig}
\usepackage{xcolor}
\usepackage{bm}
\usepackage[capitalise]{cleveref}
\newtheorem{theorem}{Theorem}[section]
\newtheorem{example}{Example}[theorem]
\newtheorem{definition}{Definition}[theorem]

\newtheorem{algorithm}{Algorithm}[theorem]
\usepackage{float}
\usepackage{graphicx,epstopdf,subfigure}
\newcommand{\as}[1]{[\textcolor{purple} {AS: {#1}}]}

\usepackage{hyperref}
\usepackage{gensymb}
\usepackage{enumitem}
\usepackage{indentfirst}
\usepackage{mathtools}
\newcommand{\ts}{\textsuperscript}

\begin{document}
\title{Decoupled synchronized states in networks of linearly coupled limit cycle oscillators}
\author{Anastasiya Salova}
\email[]{avsalova@ucdavis.edu}
\affiliation{Department of Physics and Astronomy, University of California, Davis, CA 95616, USA}
\affiliation{Complexity Sciences Center, University of California, Davis, CA 95616, USA}
\author{Raissa M. D'Souza}
\affiliation{Complexity Sciences Center, University of California, Davis, CA 95616, USA}
\affiliation{
Department of Computer Science and
Department of Mechanical and Aerospace Engineering, University of California, Davis, CA 95616, USA}
\affiliation{Santa Fe Institute, Santa Fe, NM 87501, USA}
\date{\today}	

	\begin{abstract}
	        Networks of limit cycle oscillators can show intricate patterns of synchronization such as splay states and cluster synchronization. 
        Here we analyze dynamical states that display a continuum of seemingly independent splay clusters. Each splay cluster is a block splay state consisting of sub-clusters of fully synchronized nodes with uniform amplitudes. Phases of nodes within a splay cluster are equally spaced, but nodes in different splay clusters have an arbitrary phase difference that can be fixed or evolve linearly in time.  
        Such coexisting splay clusters form a decoupled state in that the dynamical equations become effectively decoupled between oscillators that can be physically coupled.
        We provide the conditions that allow the existence of particular decoupled states by using the eigendecomposition of the coupling matrix. Additionally, we provide an algorithm to search for admissible decoupled states using the external equitable partition and orbital partition considerations combined with symmetry groupoid formalism. 
        Unlike previous studies, our approach is applicable when existence does not follow from symmetries alone and also illustrates the differences between adjacency and Laplacian coupling. 
        We show that the decoupled state can be linearly stable for a substantial range of parameters using a simple eight-node cube network and its modifications as an example.
        We also demonstrate how the linear stability analysis of decoupled states can be simplified by taking into account the symmetries of the Jacobian matrix. 
        Some network structures can support multiple decoupled patterns. 
        To illustrate that, we show the variety of qualitatively different decoupled states that can arise on two-dimensional square and hexagonal lattices.    
               
	\end{abstract}
\maketitle
		
		\section{Introduction}
		Networks of oscillators are pervasive in the world around us, from electric power grids to brain networks. 	Thus understanding the coordinated, dynamical patterns of oscillation that can spontaneously arise in such networks is of broad interest. 	Coupled oscillators provide a useful model to study many biological \cite{ashwin2016mathematical,hannay2018macroscopic,collins1993coupled,strogatz1993coupled} and engineered \cite{nishikawa2015comparative,rohden2012self} systems.
Full synchronization, where each node in the network follows the exact same trajectory in phase space, is the most basic form and widely observed~\cite{strogatz2000kuramoto, kuramoto2003chemical,ermentrout1991adaptive}. By considering the details of the network structure and dynamics, more intricate forms of synchronization can be predicted, such as cluster synchronization~\cite{belykh2008cluster,pecora2014cluster,schaub2016graph, sorrentino2016complete, golubitsky2012singularities,menara2019stability},  splay states~\cite{strogatz1993splay, zou2009splay}, chimera states~\cite{ashwin2015weak,cho2017stable,zakharova2014chimera,Abrams_2004}
and fully asynchronous states. 
Here we focus on intriguing states of synchronization  
that have been largely unstudied.  The states have intricate synchronization patterns of seemingly independent (but interwoven) sub-clusters that arise because the equations of motion lead to the cancelling out of terms of often physically coupled oscillators. Hence such a state was called ``decoupled" when first discovered~\cite{alexander1988global, alexander1986global}. Such a state appears naturally in analysis of symmetric networks of phase oscillators with homogeneous parameters \cite{ashwin1992dynamics, brown2003globally}, and was only recently achieved in experiment for a ring of phase-amplitude oscillators~\cite{matheny2019exotic}, demonstrating emergent long-range order that is a consequence of decoupling. But the range of decoupled states that can be supported on any arbitrary network topology has not yet been addressed, likewise the stability properties of decoupled states are largely unexplored.  
		
			Here we focus on decoupled states in networks of linearly coupled rotationally symmetric limit cycle oscillators (e.g. Stuart-Landau oscillators).
		Their phase shift symmetry combined with linear coupling provides an opportunity for diverse decoupled states to exist. 
		Previously, the decoupled state has been analyzed from the symmetry perspective \cite{alexander1988global,ashwin1992dynamics}.
		Here we show that the sufficient conditions on the network topology that allow decoupling can not be derived from the symmetries alone, echoing the results from groupoid formalism \cite{stewart2003symmetry,golubitsky2005synchrony} and recent cluster synchronization literature \cite{siddique2018symmetry, sorrentino2016complete, schaub2016graph}.
		This allows us to create an iterative algorithm to obtain allowed  patterns of decoupling from network structure in the distinct cases of adjacency and Laplacian coupling. 
        We use the notions of (external) equitable partitions \cite{schaub2016graph} and orbital partitions \cite{pecora2014cluster,cho2017stable} which together take into account the balanced equivalence relations of the network as well as the symmetries of the associated quotient network. All nodes within each cell of the equitable partition are fully synchronized,
        and detailed patterns of phase shift synchronization can be obtained from the symmetries of the quotient network \cite{stewart2003symmetry}.	
        	
		Additionally, we show how to use the eigendecomposition of the coupling matrix to check for admissibility of decoupled states of a given structure by generalizing recent analysis of partial synchronization in Stuart-Landau oscillator networks \cite{poel2015partial,krishnagopal2017synchronization}.
		Finally, we provide a general outline for determining the stability of these states and show how to use symmetry considerations to simplify the stability calculations using the symmetries arising from the automorphism group of the coupling matrix \cite{pecora2014cluster} and beyond \cite{golubitsky2003symmetry, emenheiser2019decoupled}. As an illustration, we explicitly perform stability calculations for a decoupled state consisting of two independent splay clusters that occurs in an example network of eight oscillators coupled on a cube (which corresponds to the case of decoupling that can be explained by symmetries alone) in cases of adjacency and Laplacian coupling. Additionally, we perform stability calculations for the same state
		for two distinct coupling topologies that are similar to the cube, but break the symmetry in ways that keeps the state admissible only for adjacency and Laplacian coupling respectively.  

		The rest of the manuscript is organized as follows. First we discuss decoupled states in more detail.  Then,
		in \cref{sec: background}, we present the necessary background, including the types of dynamics and coupling matrices we consider, the formal definition of decoupling, and the notation that will be used throughout the manuscript. 
		In \cref{sec: decoupled topology}, we consider how decoupling in linearly coupled networks arises from the network topology, expanding existing results to cases when the presence of these states is not purely dictated by symmetries. 
		To illustrate the methods, we present examples on various networks, from simple modifications of ring topology to periodic square and hexagonal lattices.
		We then illustrate how the stability calculation can be simplified based on symmetry considerations beyond cluster synchronization in \cref{sec: stability}.
		Finally, we summarize our findings and point out future directions in \cref{sec: discussion}.

 \section{Decoupled states}
		
			\begin{figure}
				\includegraphics[scale=.7]{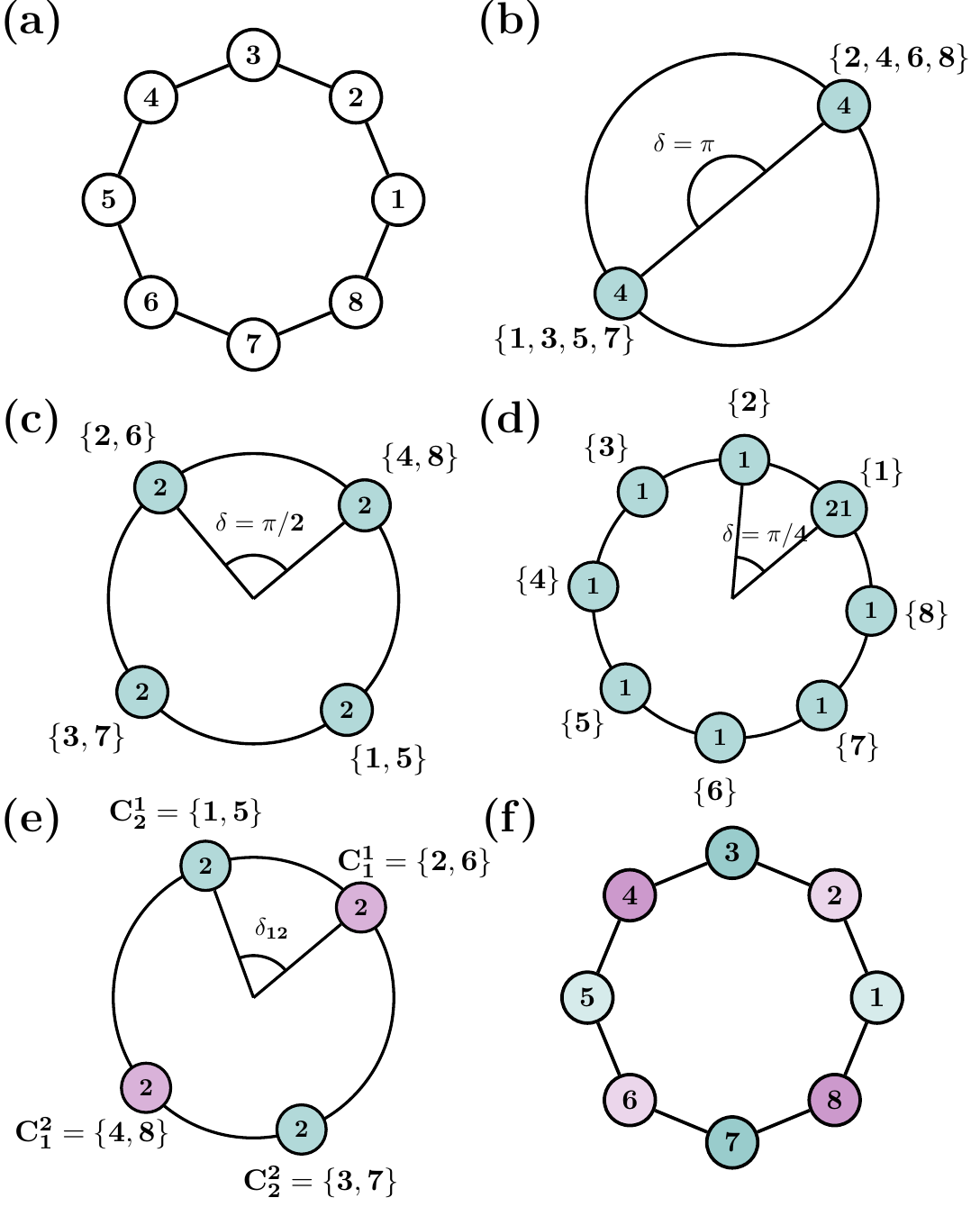}
				\caption{Illustrating splay states (b-d) and a decoupled state (e-f) in a ring network of 8 limit cycle oscillators. 
				(a) The physical coupling and node indexes. 
				(b-e)
				Each state is visualized on a ring corresponding to unit amplitude.
				Each small circle corresponds to the instantaneous phase of a subset of nodes, with the number of oscillators in that phase denoted by the number inside the circle, and their indexes provided next to the circle.  
				(b) A splay state with 2 fully synchronized sub-clusters, each consisting of 4 nodes. (c) A splay state with 4 fully synchronized sub-clusters, each consisting of 2 nodes.
				(d) A splay state with 8 fully synchronized sub-clusters, each consisting of a single node.
				(e) A decoupled state consisting of two splay state clusters (violet and teal nodes) with an arbitrary phase difference $\delta_{12}$ between them. 
				Each phase is labeled by $C_p^q=\{i,j\}$
				where $p$ denotes which \textit{splay cluster} nodes $i$ and $j$ belong to, and $q$ further partitions each splay cluster into \textit{fully synchronized sub-clusters}, each consisting of 2 oscillators. 
				(f) Visualizing the decoupled state (e) on a ring (a), where colors correspond to the splay clusters, and different intensities distinguish between different fully synchronized sub-clusters.}
				\label{fig: clusters1}
			\end{figure}

		A decoupled state consists of several distinct splay clusters. Each splay cluster is a block splay state,
		and nodes in different splay clusters have an arbitrary phase difference that can either be fixed in time, meaning the state is periodic, or evolve linearly for quasiperiodic states. Here, the term \textit{decoupled} refers to the fact that the coupling terms responsible for intra-cluster interactions cancel out in the dynamical equations, even if physical coupling between the oscillators in different splay clusters is present. 			
		Block splay states (also referred to as twisted states or traveling wave states in the literature~\cite{strogatz1993splay, zou2009splay}) arise as a form of symmetry breaking for a variety of coupling matrix structures, for instance, rings \cite{emenheiser2016patterns, matheny2019exotic, wiley2006size}, all-to-all coupling \cite{ku2015dynamical}, and lattices of oscillators \cite{hakim1992dynamics, lee2018twisted}. 
		These states are characterized by synchronization with a nonzero winding number (meaning that for some ordering of oscillators, each pair of neighbors is separated by an equal phase difference, and the winding number is determined by how many of the differences add up to $2\pi$). See, for instance, Fig.~\ref{fig: clusters1}(b-d) which show three distinct block splay states that can exist for a ring of eight identical oscillators.  Such a ring supports a decoupled state consisting of two independent splay clusters as shown in Fig.~\ref{fig: clusters1}(e), with $\delta_{12}$ denoting the arbitrary phase difference between the two splay clusters. 
The oscillators can be simple phase oscillators or can be
		limit cycle oscillators, such as nanoelectromechanical oscillators~\cite{emenheiser2016patterns, matheny2019exotic} and Stuart-Landau oscillators, where further patterns including
amplitude death can exist~\cite{poel2015partial, krishnagopal2017synchronization}. 
		
		A periodic decoupled state was first observed in an analysis of bifurcations on rings of coupled oscillators \cite{alexander1986global}, and motivated a symmetry based analysis of decoupling phenomena in networks with more general structure \cite{alexander1988global}.
		Under the presence of symmetries in nodal dynamics and with linear coupling, these states can be present in a diverse range of systems such as oscillators coupled in rings, hypercubes, full bipartite graphs, and infinite chains \cite{alexander1988global}.  
		The state appears naturally in analysis of symmetric networks of phase oscillators with homogeneous parameters \cite{ashwin1992dynamics, brown2003globally} (the latter considers a general case of difference and product coupling), where the state is described as \textit{partitioned rotating blocks}. 
		
		More recently, the state was predicted to occur in rings of $4n$ coupled phase-amplitude nanoelectromechanical oscillators \cite{emenheiser2016patterns}, and, to our knowledge, experimentally observed for the first time in 2019 \cite{matheny2019exotic}. 
		Interestingly, in ring networks, the physically coupled neighboring nodes have an arbitrary phase difference and are effectively decoupled from each other.
		Stability analysis of the decoupled states on $4n$-node ring networks with uniform and non-uniform natural frequencies was performed in Ref. \cite{emenheiser2019decoupled}. 
		There, it was shown that the amplitude degree of freedom is essential for the linear stability of that state (i.e., the state would not be linearly stable for phase only oscillators in the absence of explicit long-range, nonpairwise, or nonphase coupling).
		Moreover, it was demonstrated that mismatches in the natural frequencies of oscillators
		lead to intricate patterns of stability that show sensitive dependence on the relative frequencies of groups of oscillators in distinct splay clusters.

		\section{Formalism} \label{sec: background}
		
		In this manuscript, we consider networks of phase shift invariant limit cycle oscillators (such as Stuart-Landau oscillators) with linear coupling, with dynamical behavior governed by: 
		\begin{align}\label{eq: general form}
		\dot{z}_j=f(|z_j|, \psi_j)\cdot z_j+\kappa_{jk}e^{i\beta_{jk}} \sum\limits_k M_{jk}z_k.
		\end{align}
		The first term corresponds to the evolution of the state of each oscillator, 
		denoted by, $z_j=r_j e^{i\theta_j}\in \mathbb{C}$, in absence of coupling. 
		Without coupling, the system would evolve according to a nonlinear function $f(|z_j|,\psi_j)\cdot z_j$ (where the nonlinearity arises from the form of the function $f$). Specifically, we consider functions where, in absence of coupling, one of the admissible states of the system is a limit cycle with fixed amplitude $r_j$ and linearly evolving phase $\theta_j$.
		Here, $\psi_j$ denotes the parameters of individual oscillators.
	 
		The oscillators are coupled through the coupling matrix $M$, and parameters $\kappa_{jk}$ and $\beta_{jk}$ contribute to the strength of dissipative coupling $\text{Re}(\kappa_{jk}e^{i\beta_{jk}})$ and reactive coupling $\text{Im}(\kappa_{jk}e^{i\beta_{jk}})$.
		The coupling matrix $M$ is constrained to have binary off-diagonal entries and can either correspond to the adjacency matrix $A$ or the graph Laplacian $L$, where $L=A-D$ (note the sign convention) 
		the diagonal matrix $D$ is defined by $D_{ij}=\delta_{ij}\sum A_{i}$, and $\delta_{ij}$ is the Kronecker delta. Here, we assume the network is undirected. However, the results generalize for directed networks as well.	
		
		\iffalse The analysis of decoupled states has been mostly absent from literature beyond phase oscillators \cite{brown2003globally, ashwin1992dynamics} and symmetry considerations \cite{alexander1988global}.  
		\fi
		
		States of dynamical systems described by \cref{eq: general form} and its modifications (e.g., with added time delay) have been extensively analyzed in case of Stuart-Landau oscillators, where the individual oscillator dynamics are of the form: 
		\begin{align}\label{eq: sl one}
		f(|z_j|,\psi_j)\cdot z_j=(\lambda_j+i\omega_j -|z_j|^2)z_j,
		\end{align}
		where $\lambda_j \in \mathbb{R}$ is the bifurcation parameter, and $\omega_j \in \mathbb{R}$ is the individual oscillator frequency.
		Similarly, the dynamics of experimentally realized nanoelectromechanical oscillators \cite{matheny2019exotic} is well approximated by equations of the form of \cref{eq: general form} \cite{lifshitz2008nonlinear}, with individual oscillator dynamics following:
		\begin{align}\label{eq: nems one}
		f(|z_j|,\psi_j)\cdot z_j=\left(-1 + i\omega_j + 2i\alpha_j|z_j|^2 + \dfrac{1}{|z_j|}\right)z_j,
		\end{align}
		where $\omega_j\in \mathbb{R}$ is the oscillator frequency, and $\alpha_j$ is the Duffing nonlinearity.
		We will use these types of individual oscillator dynamics in our illustrative examples throughout the manuscript.
		
		Some basic parameters are needed to define a state consisting of multiple splay clusters. We illustrate this with an example of a decoupled state in a network of nanoelectromechanical oscillators shown in \cref{fig: clusters1} (e-f) and studied in Ref.\cite{emenheiser2019decoupled}. 
		Let $k$ denote the number of independent splay clusters. Let $m$ denote the number of fully synchronized sub-clusters in the splay cluster, and let $n$ denote the number of nodes in each sub-cluster.  The number of nodes in the network is simply $N=kmn$. For \cref{fig: clusters1} (e-f), $k=2, m=2, n=2$.   We will also use the key notation $C^q_p$ with $p \in {1, 2, \cdots k}$ indicating the splay cluster index and $q \in {1,2, \cdots m}$ indicating the fully synchronized sub-cluster index. We provide formal details below (including the more complicated case where $m$ and $n$ can vary for different splay clusters).
		
		\begin{example}\label{nems1}
			A network of eight nanoelectromechanical oscillators coupled to their nearest neighbors via Laplacian coupling on a ring (shown schematically on \cref{fig: clusters1} (a)) exhibits a variety of states that can be observed in experiment \cite{matheny2019exotic}.
			Assuming the parameters of the network are homogeneous, the equation below is a good approximation of the dynamics of the system: \begin{align}\label{eq:complexphase}
			\begin{gathered}
			\dot{z_j} = -z_j + i\omega z_j + 2i\alpha|z_j|^2z_j + \frac{z_j}{|z_j|}
			\\ + i\delta\sum\limits_{k=j\pm 1}A_{jk}\left(z_k-z_j\right).
			\end{gathered}
			\end{align}
			In addition to multiple splay states, some of which shown on \cref{fig: clusters1}(b-d),
			this system admits a decoupled state (shown in \cref{fig: clusters1} (e-f)) defined by: 
			\begin{align}
			(z_1,...,z_8)=(z_1,z_2,-z_1,-z_2,z_1,z_2,-z_1,-z_2).
			\end{align} 
			Here, $z_1=e^{i\theta_1}$, $z_2=e^{i\theta_2}$, and $\theta_2-\theta_1$ is an arbitrary free parameter, and the oscillators are labeled by going around the ring.
			
			Using the $C_p^q$ notation (defined formally below in \cref{def: decoupled state}), $C_1^1=\{1,5\}$, $C_1^2=\{3,7\}$, $C_2^1=\{2,6\}$, $C_2^2=\{4,8\}$, and $C_1=\{1,3,5,7\}$, $C_2=\{2,4,6,8\}$. So the ``even'' nodes are decoupled from the ``odd'' nodes (evident from neighbors of each node being in antiphase in relation to each other), and the neighboring nodes that are physically coupled are separated by an arbitrary but fixed phase difference. 
			
			If the parameters in \cref{eq:complexphase} are homogeneous, all the nodes (and therefore the splay clusters) move with the same frequency, resulting in the constant time-independent phase difference $\theta_2-\theta_1$. 
		\end{example}
		 
		We now define the (periodic or quasiperiodic) decoupled state more generally as a combination of different splay clusters, where each splay cluster can have a unique number of fully synchronized sub-clusters as well as nodes within each sub-cluster:
		\begin{definition}\label{def: twisted state}
			We say a set of $mn$ nodes is in a \textbf{splay state} (with $m$ fully synchronized clusters) if for some ordering of nodes their states are $\{z,...,z,\omega z,...,\omega z ..., \omega^{m-1}z,..., \omega^{m-1}z\}$. 
			Here, $\omega$ is the $m^{\text{th}}$ primitive root of unity.
		\end{definition}
		\begin{definition} \label{def: decoupled state}
			Here we define a \textbf{decoupled state} consisting of $k$ distinct but interleaved splay states, where each splay state is called a \textbf{splay cluster} for conceptual convenience. 
			
			Let the nodes of the system be labeled by indexes $I=\{1,...,N\}$, where $N$ is the total number of nodes. 
			We can partition the nodes according to their phases into non-overlapping clusters $C_p^q$ with \textbf{subscripts} indicating the \textbf{splay cluster} to which the node belongs 
			and the \textbf{superscripts} indicating the \textbf{fully synchronized sub-cluster} within each splay cluster. 
			Let the splay state clusters be indexed by $p$. Then $m_p$ is the number of fully synchronized sub-clusters in splay cluster $p$ and 
			%that state consists of, and 
			$n_p$ is the number of nodes in each of the fully synchronized sub-clusters (the total number of nodes is then $N=\sum\limits_{p=1}^k n_p m_p$.).
			
			Mathematically, this is described by:
			\begin{itemize}   \setlength\itemsep{0em}
				\item $C_p^q$, $p=1,...,k$, $q=1,...,m$, s.t. if $i,j\in C_p^q$, $z_i=z_j$, and $|C_p^q|=n_p m_p$;
				\item $C_p=C_p^{q_1}\cup ... \cup C_p^{{q_m}_p}$, s.t. if $i\in C_p^{q_r}$ and $j\in C_p^{q_s}$, $z_i=e^{2\pi i |q_r-q_s|/m} z_j$, and the phase difference is fixed over time;
				\item if $i\in C_{t}^{q_1}$ and $j\in C_{u}^{q_1}$, $z_i=e^{i \delta_{tu}}\dfrac{r_i}{r_j} z_j$, and the condition holds instantaneously, but $\delta_{tu}$ is allowed to evolve linearly in time.
			\end{itemize}
			
			The partition of nodes into the cells $C_1^1,...,C_1^{m_1},...,C_k^1,...,C_k^{m_k}$ defines the state.
			
			In presence of adjacency coupling, the \textbf{decoupling} is manifested by the fact that the total effect of the nodes in the cluster $C_{p}$ on each node $j$ in the cluster $C_{r}$ ($p\neq r$) cancel out. 
			Mathematically, the interaction terms are proportional to $\sum\limits_{k\in C_{p}}A_{jk}z_k=0$. In case of Laplacian coupling, the only effects of the nodes in $C_{p}$ on each node $j$ in the cluster $C_{r}$ ($p\neq r$) are manifested through the self-interaction terms. Mathematically, the interaction term affecting the state of the node $j$ if proportional to $\sum\limits_{k\in C_{p}}A_{jk}(z_k-z_j)=-n_{pj} z_j$, where $n_{pj}$ denotes the number of edges coming into the node $j$ from the cluster $p$.
		\end{definition}
		
\begin{figure}
			\includegraphics[scale=.6]{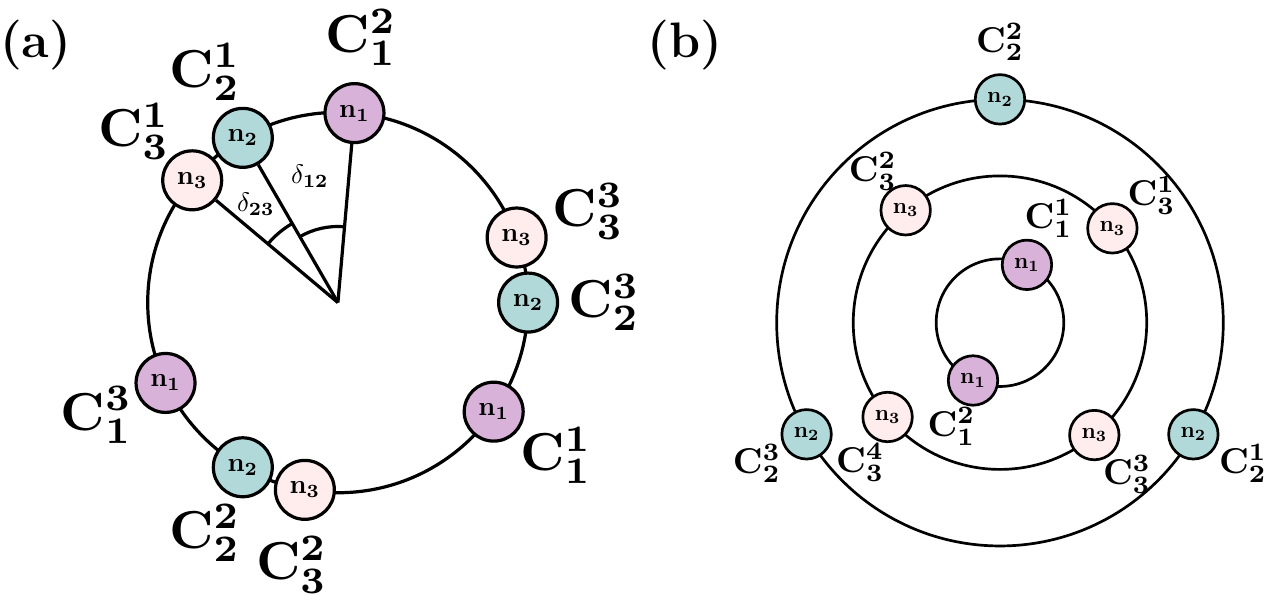}
			\caption{Schematic examples of possible phase (position on a ring) and amplitude (ring radius) configurations in decoupled states in networks of limit cycle oscillators. 
				Different colors represent different splay clusters. $C_p^q$ corresponds to $q$\ts{th} fully synchronized sub-cluster in $p$\ts{th} splay cluster.
				(a) Equal amplitude state with constant phase differences, $m_1=m_2=m_3=3$ (could be periodic or quasiperiodic, depending on the details of the system).
				(b) Quasiperiodic state with different amplitudes, $m_1=2,~~ m_2=3,~~m_3=4$.
			}
			\label{fig: clusters}
		\end{figure}

		To further illustrate how the oscillators are organized into splay clusters and fully synchronized sub-clusters in a decoupled state, and how this corresponds to our notation, we present schematic examples of possible amplitudes and relative phases of such states in \cref{fig: clusters} for phase-amplitude oscillators. 
		\cref{fig: clusters}(a) corresponds to the case when the amplitudes of all the oscillators are equal.
		In contrast, \cref{fig: clusters}(b) shows a state where the amplitudes of oscillators in different splay clusters differ. Moreover, each splay cluster has a distinct number (2, 3, and 4) of fully synchronized sub-clusters, which leads to interesting multi-frequency oscillation behavior. 
		
		A state where both antiphase synchronization and multifrequency behavior are present, similarly to the one on  \cref{fig: clusters}(b), is described in Ref.\cite{tumash2019synchronization}. However, in the state presented there, clusters of oscillators of different amplitudes all either have the same or opposite phases, making it different from the decoupled state considered in our manuscript.
			
		We note that the dynamics described by \cref{eq: general form} is not the only type of dynamics producing decoupled clusters.
		For example, the state can arise in mean field coupled networks of phase-only oscillators such as Kuramoto and Kuramoto-Sakaguchi oscillators \cite{brown2003globally, ashwin1992dynamics}, which can be considered as the approximation linearly coupled Stuart Landau oscillators in case of weak coupling. 
		Such models have fewer degrees of freedom and may be easier to analyze, but do not capture the amplitude dynamics or full stability properties of the decoupled states of phase-amplitude oscillators \cite{emenheiser2019decoupled}.
	
	    In this section, we presented the general form of a decoupled state. 
	    Now, a natural question to ask is what coupling topologies admit its existence.
	    In addition to the all-to-all case, some examples of these networks have been investigated in literature, e.g., in context of rings of Josephson junctions \cite{alexander1986global} and nanoelectromechanical oscillators \cite{matheny2019exotic}.
	    More generally, in \cite{alexander1988global}, the constraints on the topologies allowing the existence of decoupled states are formulated based on symmetry considerations for a more general class of dynamical equations with adjacency coupling. 
		In \cref{sec: decoupled topology} of this manuscript, we demonstrate that the decoupled states do not uniquely arise from symmetries alone, even in case of adjacency coupling.
		Therefore, our analysis covers cases not addressed in earlier work.
		
		\section{Decoupled states and network topology} \label{sec: decoupled topology}
		
		In this section, we investigate what decoupled states are admissible for a given network topology and, conversely, what network topologies are allowed to form a decoupled state of a given form. Linearity of the dynamics in $z$ allows directly predicting the form of allowed decoupled states from the eigendecomposition of the adjacency or Laplacian matrix of the network, as we show in \cref{subsec: eigenvectors}. Checking the admissibility of a given form of the decoupled state is based on these conditions is trivial. However, finding all the allowed decoupled states based on the conditions in \cref{subsec: eigenvectors} hold can be challenging in practice, as the corresponding eigenvalues can be highly degenerate.
		
		An alternative way to search for decoupled states is by investigating the structure of the coupling network directly. Previously, the decoupling conditions were formulated for oscillator networks with adjacency coupling and symmetries \cite{alexander1988global}. However, we show that decoupling is admissible for a much wider range of network topologies. We build upon work showing that cells of equitable partition of networks can synchronize \cite{golubitsky2005synchrony, schaub2016graph, siddique2018symmetry, o2013observability}, and more intricate patterns of synchrony can be inferred from the symmetries of the quotient networks \cite{stewart2003symmetry,golubitsky2016rigid}. Combining these results with adjacency and Laplacian coupling and taking the decoupling effects into account, we formulate the conditions on decoupling that expand the set of networks previously discussed in \cref{subsec: algorithm}. In both cases, the network parameters can have modular structure, as discussed in \cref{subsec: heterogeneous}. \cref{section: lattices} demonstrates how the algorithms in \cref{subsec: algorithm} can be used to reveal ways in which the decoupled state and its combination with amplitude death can appear in coupled oscillators on 2D square and hexagonal lattices.

	It is worth noting that the analysis below can be extended to include the concept of amplitude death, 
	which is a phenomenon associated with stabilization of the trivial steady state solution that can be observed in Stuart Landau oscillator networks \cite{mirollo1990amplitude}.
		For simplicity, most of the results below are presented for the state in which all the oscillators have nonzero amplitudes, and thus no partial amplitude death \cite{atay2003total} is observed.
		However, adding nodes exhibiting amplitude death to the adjacency coupled network in a way that keeps the dead nodes decoupled from other nodes would not change the admissibility of state of the other nodes (though it will influence its stability properties). 
This also extends to Laplacian coupling. 		
		Specifically, nodes that are only connected to one fully synchronized cluster can be synchronized with the rest of the cluster without destroying the decoupled state on other nodes.
		
		\subsection{Admissible patterns of decoupling from eigendecomposition}\label{subsec: eigenvectors}
		In this section, we discuss the conditions on admissibility of a specific decoupled state given the coupling via an adjacency/Laplacian matrix $M$. 
		These conditions can be formulated in terms of the eigenvectors of $M$ (similar to Ref.\cite{poel2015partial,krishnagopal2017synchronization}, where the concept of eigensolutions is discussed).
		Assuming the parameters are homogeneous throughout the network, the dynamics of the system in \cref{eq: general form} reduces to:
		\begin{align}\label{eq: general form matrix}
		\dot{z}=f(|z|)\cdot z+\kappa e^{i\beta} Mz.
		\end{align}
		The coupling matrix can be decomposed according to:
		\begin{align}
		Mv=\eta v,
		\end{align}
		where $\eta$ and $v$ are its eigenvalues and eigenvectors.
		
		First, we seek eigendecompositions that result in a \textit{periodic} decoupled state that evolves according to:
		\begin{align}\label{eq: ansatz}
		z(t) = z_0 r_{\eta} e^{i\omega_{\eta}t},
		\end{align}
		where $z_0$ is an initial condition corresponding to the decoupled state as defined in \cref{def: decoupled state}. 
		
		\cref{eq: ansatz} suggests that the conditions on the eigendecomposition of the coupling matrix $M$ (again, $M=A$ for adjacency coupling, and $M=L=A-D$ for Laplacian coupling) that result in a periodic decoupled state with $z_0$ are:
		\begin{itemize}\setlength\itemsep{0em}
			\item If $i,j\in C_p^{q}$: 
			\begin{align}\label{eq: bp 1}
			[v_{p}]_i=[v_{p}]_j.
			\end{align}
			\item If $i\in C_p^{q_1}$ and $j\in C_p^{q_2}$:
			\begin{align}\label{eq: bp 2}
			[v_{p}]_j=e^{2\pi i|q_1-q_2|/m}[v_{p}]_i.
			\end{align}
			\item If $i\notin C_{p}$: 
			\begin{align}\label{eq: bp 3}
			[v_{p}]_i=0.
			\end{align}
			\item For distinct clusters $C_{p_1}$ and $C_{p_2}$:
			\begin{align}\label{eq: bp 4}
			\eta_{p_1}=\eta_{p_2},
			\end{align}
			which results in any vector corresponding to the decoupling state $v_{\text{d}} = \sum e^{i\delta_{j}} v_{p_j}$ being an admissible eigenvector. 
		\end{itemize}
        To summarize, we require that any vector with the entries corresponding to an instance of a decoupled state with an arbitrary phase difference between splay clusters is an eigenvector of $M$.
	
		We can now show that these conditions on the eigendecomposition lead to decoupling. Let $v_d$ be an eigenvector. We use an ansatz $z_i=[v_d]_i z_\eta$ to obtain:
	\begin{align*}
	v_d\circ\dot{z}_{\eta}=\left(f(|v_d\circ z_{\eta}|)+\kappa e^{i\beta}M\right) v_d \circ z_{\eta},
	\end{align*}
	where $\circ$ denotes element-wise product.
		Using the rotational symmetry of the system, we arrive to:
		\begin{align}
	 	\dot{z}_{\eta} = \left(f(|{z}_{\eta}|)+\eta\kappa e^{i\beta}\right) z_{\eta}
		\end{align}
		The resulting solution is then of the form:
		\begin{align}\label{eq: anz}
		z(t) = r_{\eta}e^{i\omega_{\eta} t} v_d,
		\end{align}
		which corresponds to the evolution of a decoupled state.
		The amplitude $r_{\eta}$ and the angular frequency $\omega_{\eta}$ depend on the parameters and the form of the function $f$.	

		A condition similar to \cref{eq: bp 4} (normalized $v_{p_1}+v_{p_2}$ is an eigenvector of the Laplacian) without \cref{eq: bp 1,eq: bp 2,eq: bp 3} holding does not result in a decoupled state. For instance, an adjacency/Laplacian matrix for a ring of six coupled oscillators has eigenvectors corresponding to splay clusters (rotating blocks) with winding numbers 2 and 3, but these states are not decoupled. %On the other hand, 
Yet, multiple decoupled states are admissible for all-to-all coupled networks of six oscillators, as shown in \cref{fig: six nodes} (c-d).
		
		On the other hand, if all the conditions in \cref{eq: bp 1,eq: bp 2,eq: bp 3} hold, but \cref{eq: bp 4} does not hold, the resulting decoupled state is quasiperiodic. That quasiperiodic state is characterized by:
		\begin{align}
		\begin{gathered} \label{eq: multifreq}
		z_{j} = z_{\eta, p} [v_p]_j,\\ 
        z_{\eta, p}= r_{\eta,p}e^{i\omega_{\eta,p} t}, 
        \end{gathered}
		\end{align}
		for $j\in C_p$. This allows the oscillator amplitudes and frequencies, $r_{\eta, p}$  and $\omega_{\eta, p}$, to be different for nodes in different splay clusters, as shown in \cref{fig: clusters} (b).
		
	    Our approach extends the analysis of eigensolutions describing the dynamics of constant amplitude states in networks of coupled Stuart-Landau oscillators, as presented in Refs. \cite{poel2015partial,krishnagopal2017synchronization}. 
		There, the eigendecomposition was associated with splay states and their coexistence with amplitude death (the state in which $z(t)=0$ for some oscillators) \cite{mirollo1990amplitude}, but decoupling was not considered. We discuss the case of Stuart Landau oscillators in more detail in \cref{app: SL}. 
		For networks of Stuart Landau oscillators, the decoupled state can coexist with partial amplitude death. For instance, if the decoupled state is an admissible state, the state a set of nodes belonging to a decoupled cluster has zero amplitude is admissible as well. Additionally, \cref{eq: multifreq} demonstrates that multifrequency oscillations previously associated with coexistence with amplitude death can occur even in absence of it.
		\begin{example}\label{ex: nems}
			As described in \cref{nems1}, a decoupled state exists in an eight oscillator ring with nearest neighbor coupling, and the solution is valid both for adjacency and Laplacian coupling schemes. 
			This is evident since the following eigenvectors that share an eigenvalue correspond to decoupling of the even and odd nodes on the ring:
			\begin{align}
			\begin{gathered}
			v_{p_1}=(1,0,-1,0,1,0,-1,0)^T,\\
			v_{p_2}=(0,1,0,-1,0,1,0,-1)^T,\\
			\eta_{p_1}=\eta_{p_2}.
			\end{gathered}
			\end{align}
			In case of Laplacian coupling, breaking the symmetry by adding an edge between any pairs of nodes related by $i,j \in C_p^q$ does not affect these eigenvectors, and the state is still admissible. However, this does not hold for adjacency coupling.	
		\end{example}
	    
	    We also note that this if the state is a result of symmetries (when the state belongs to a fixed point subspace of an isotropy subgroup of the symmetry group of the system \cite{golubitsky2012singularities}), these conditions \textcolor{black}{become identical to the ones presented in Ref. \cite{alexander1988global}} and could result in periodic or quasiperiodic states \cite{emenheiser2019decoupled}.
	    
	    Using the approach outlined here is easy if the goal is to check whether a particular decoupled state is admissible for a coupling topology encoded in $M$. However, it is not always trivial to enumerate the decoupled states of larger networks based on the eigendecomposition of the coupling matrix. This method is not very easily applicable to highly regular networks, such as periodic lattices discussed in \cref{section: lattices}, because the eigenvalues become highly degenerate, making it nontrivial to check conditions in \cref{eq: bp 1,eq: bp 2,eq: bp 3,eq: bp 4} numerically. In the following section, we introduce an alternative method that can be used to address this issue and allows to make a more intuitive connection between the network structure and the decoupled state dynamics.
		
		\subsection{Admissible patterns of decoupling from network partitions}\label{subsec: algorithm}
		
		In this section, we provide a complementary approach of relating network structure to the existence of decoupled states. 
		Specifically, we provide an iterative algorithm that partitions the Laplacian coupled network nodes into splay clusters $C_p$ and fully synchronized clusters $C_p^q$ based on the concepts of equitable and orbital partitions used in cluster synchronization literature \cite{pecora2014cluster, pecora2017discovering, schaub2016graph, siddique2018symmetry, sorrentino2016complete}. We also show how the adjacency coupling case is different and provide  a way to search for decoupled states in that case as well.
		
		For both coupling types, we require that the coupling between the nodes within a splay cluster admits a splay state. However, in the Laplacian case, the edges between any nodes in the same fully synchronized cluster can be ignored, whereas in case of adjacency clustering these edges also affect the admissibility of the state. Additionally, we have take into account that self-terms can arise from Laplacian coupling between different splay clusters, which is not an issue for adjacency case.
		
		Let the set of nodes in the network be defined by $I$. 
		\begin{definition}
			An \textbf{equitable partition} is a partition of the indexes of network nodes into non-overlapping \textbf{cells} (collections of nodes) $I_1,...,I_d$, where the number of edges from a node in $I_i$ to a node in $I_j$ is uniquely determined by the indexes $i$ and $j$. 
			An \textbf{external equitable partition} is a partition where these conditions hold for $i\neq j$.
		\end{definition}
		\begin{definition}
			An \textbf{orbital partition} is a partition of network nodes into \textbf{cells} $I_1,...,I_d$ according to the orbits of the automorphism group (the symmetry group formed by the permutations of the indexes of network nodes). The nodes that permute among one another under the action of all symmetry group elements get assigned to the same cell.
		\end{definition}
		\begin{definition}
			A \textbf{quotient network} associated with a particular a coarse-grained version of the original network, such that each cell of the that partition becomes a new node and the weights between these new nodes are the out-degrees between the cells in the original graph.
		\end{definition}
	
		Now, we can define an algorithm to obtain decoupled states in networks with Laplacian coupling. 
		(The algorithm for adjacency coupling is presented in Alg.~\ref{alg2}.)
		In contrast with existing decoupled state literature \cite{alexander1988global, ashwin1992dynamics}, we consider the states that arise from more than symmetries alone.
		
		\begin{algorithm}\label{alg1}
			To find admissible decoupled states  for \textbf{Laplacian} coupling, it is sufficient to:
			\begin{enumerate}\setlength\itemsep{0em}
				\item Find an external equitable partition of the network and form an quotient network associated with that partition (it is possible that the equitable quotient is the same as the original network).\label{s1}
				\item \label{orb}Find an orbital partition of the external equitable quotient network obtained above according to the symmetry group $Z_{m_1}\times ... \times Z_{m_k}$  and form a nontrivial quotient network associated with the orbital partition. The action of the symmetry group is constrained by $Z_{m_1},...,Z_{m_k}$ all permuting non-overlapping subsets of nodes, and by the union of these subsets of nodes being the entire network.  
				\item Check if the number of edges coming \textbf{into} every node is divisible by the degree of that node.
				\begin{itemize}\setlength\itemsep{0em}
					\item If that is \textbf{not} satisfied, go back to step 2 and try a new orbital partition. In case that does not work, go to step 1 and try a new external equitable partition.
					\item If that is satisfied, then a decoupled state is obtained. The orbital partition provides the assignment into splay clusters $C_p$, and the equitable partition provides the assignment into fully synchronized clusters $C_p^q$.
				\end{itemize}
			\end{enumerate}
		\end{algorithm}
		
		In a special case of homogeneous parameters, if $m_1=...=m_k$, and the weights on all self-loops and edges in an orbital quotient network are equal, the state is periodic.
		
		We show that these conditions
		 indeed characterize the decoupled states.
		
		The first step is forming an equitable partition. 
		It has been shown that equitable partitions lead to synchronization \cite{siddique2018symmetry, schaub2016graph}, since all the nodes in a given partition cell get the same input from all the other cells. 
		Thus, the nodes in each cell of that partition can synchronize. 
		
		\begin{figure}
			\includegraphics[scale=.7]{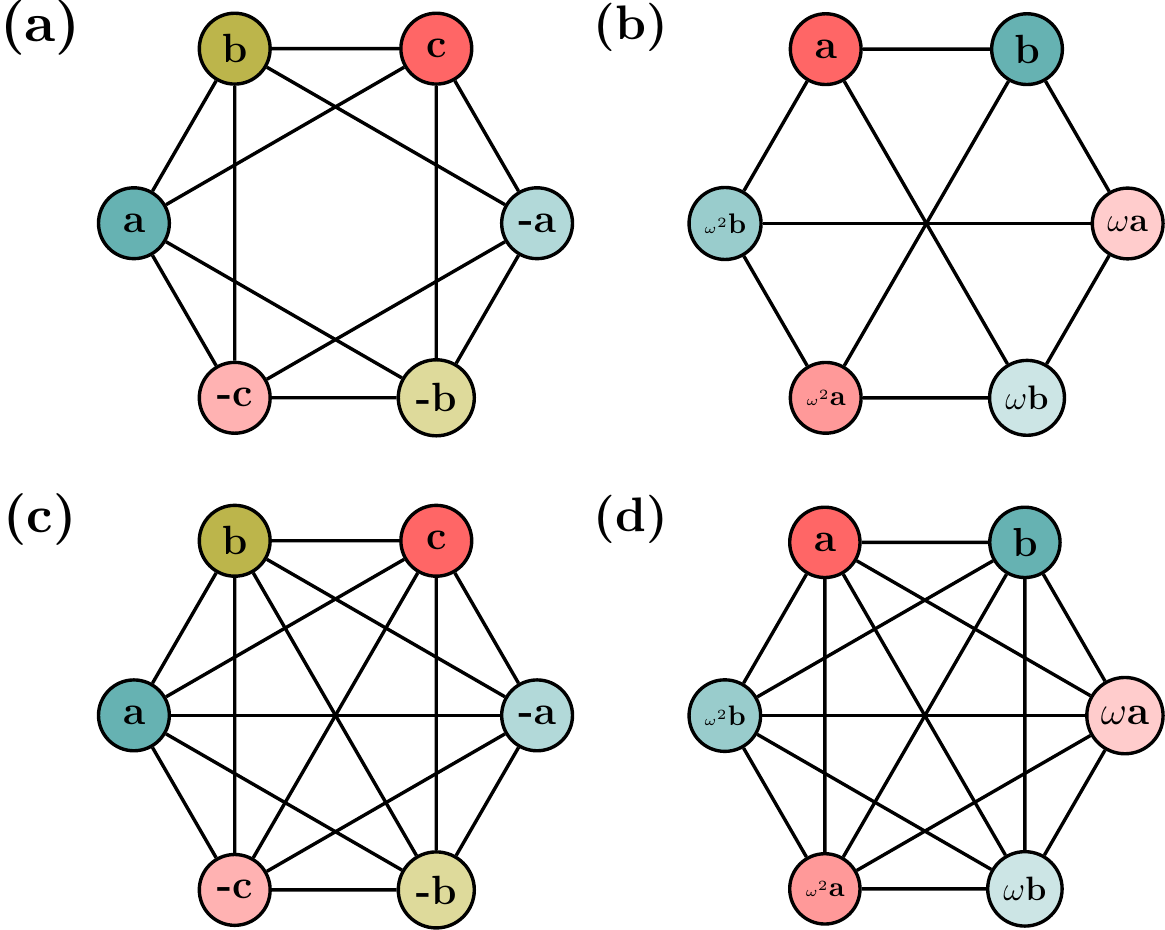}
			\caption{Examples of decoupling on a 6-node network. A decoupled state does not appear in a ring with nearest neighbor coupling.  (a) Three decoupled clusters in case of next neighbor coupling. (b) Two decoupled clusters for nearest neighbor coupling with additional edges across the ring ($\omega$ stands for $e^{2\pi i/3}$). (c-d) Both of the states are admissible for all-to-all coupling.}
			\label{fig: six nodes}
		\end{figure}

		The second step is forming an orbital partition with respect to a symmetry group $Z_{m_1}\times ... \times Z_{m_k}$, where $Z_{m_i}$ refers to the cyclic group of degree $m_i$. 
		This ensures that a block splay state exists within each cluster \cite{golubitsky2010network,golubitsky2012network}. We note that the symmetry of the quotient network does not necessarily translate to the symmetry of the original network. 
		
		The last step combined with the second step ensures that each of the nodes in a decoupled state is not influenced by other splay clusters in the dynamical equations. 

		We also note that the steps are not uniquely defined. First, there can be multiple equitable partitions of the network. 
		In some cases, different network partitions correspond to qualitatively distinct decoupled states. For instance, in a network of 6 oscillators, two distinct decoupled states are possible, see \cref{fig: six nodes}. 
		One of them corresponds to $m=3$ with $n_1=n_2=1$, the other to $m=2$ with $n_1=n_2=n_3=1$. Another case corresponds to finding partitions that could be further refined to produce decoupled states. In that case, the state obtained is a restriction of a decoupled state with more clusters to a specific inter-cluster phase difference. Finally, the same quotient network structure could correspond to qualitatively different decoupled states (e.g., shown in \cref{section: lattices}).

		Below is an example of applying the algorithm:
		
		\begin{example}\label{nems2}
			We consider a network of eight oscillators similar to the ring in \cref{nems1} with additional edges between a subset of next nearest neighbors and two opposite nodes, as shown in \cref{fig: clusters_algo}(a). 
			\begin{itemize}\setlength\itemsep{0em}
				\item Solid lines on \cref{fig: clusters_algo} represent the ring topology, which would result in a periodic decoupled state in absence of parameter heterogeneity. The state is admissible in case of adjacency or Laplacian coupling. 
				\item Adding additional edges represented by dotted lines on \cref{fig: clusters_algo} would result in a quasiperiodic state. The state is admissible in case of adjacency or Laplacian coupling. 
				\item Adding an edge represented by dashed line on \cref{fig: clusters_algo} makes that quasiperiodic state only admissible in case of Laplacian coupling.
			\end{itemize}
			
			 Following \cref{alg1}, we find the decoupled state that the network admits for the Laplacian coupling case. The steps of the algorithm, as well as their description, is shown on \cref{fig: clusters_algo}. \cref{fig: clusters_algo} (d) shows the instantaneous state of the system. In general, even if all the oscillator natural frequencies are equal, the frequencies of splay clusters (corresponding to nodes colored red and blue on \cref{fig: clusters_algo} (c)) will differ because of the different intra-cluster connectivity structure (as evident on \cref{fig: clusters_algo} (c$'$)).  This leads to multifrequency splay synchronization of identical oscillators without amplitude death.
			
			\begin{figure}
				\includegraphics[scale=.4]{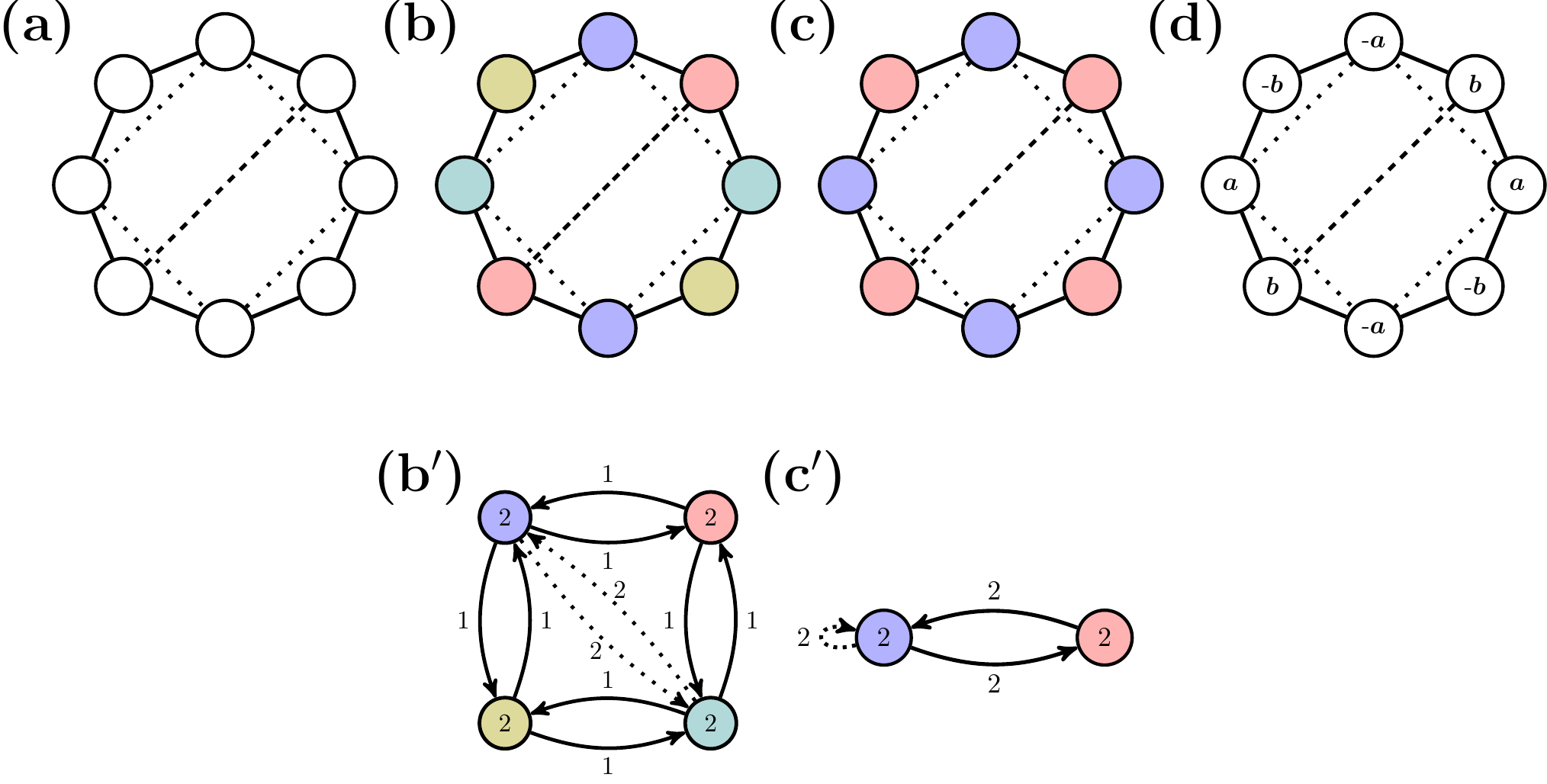}
				\caption{Obtaining the decoupled state using \cref{alg1} for the network topology shown on (a). 
				Subfigures (b)-(c) represent the steps to obtain the state. The top row shows network partitions,  and the bottom rows demonstrate the quotient networks on various steps of \cref{alg1}. 
				(b) shows the four  fully synchronized clusters corresponding to the external equitable partition with the corresponding quotient network shown on (b$'$), once the self edges are removed. 
				(c$'$) shows the quotient network with re-weighted edges that is now colored according to its orbital partition. 
				(c) shows the original network colored according to the orbital quotient. 
				The decoupled state consists of four fully synchronized clusters (shown in distinct colors on (b)) and two splay (rotating block) clusters (shown in distinct colors on (c)).
			    (d) demonstrates the form of the state, where $a$ and $b$ are complex state variables.}
				\label{fig: clusters_algo}
			\end{figure}
		\end{example}
		
		This same algorithm can be applied to periodic lattices. The analysis of decoupled states in square and hexagonal 2D periodic lattices is presented in \cref{section: lattices}. There, we show that diverse synchronization patterns can arise for the same regular network connectivity patterns.
		
		Now, we consider adjacency coupling.
		If the external equitable partition is replaced by an equitable partition in the first step of \cref{alg1}, that algorithm can be used to obtain decoupled states for adjacency coupling.
		However, that does not cover the full range of possibilities for decoupled states arising in adjacency coupled networks.
		We provide a more general algorithm below.
			\begin{algorithm}\label{alg2}
			To find admissible decoupled states  for \textbf{adjacency} coupling, it is sufficient to:
			\begin{enumerate}\setlength\itemsep{0em}
				\item Divide the network nodes into non-overlapping sets of nodes $I_1,...,I_k$. This partition is a candidate for assigning nodes into splay clusters $C_p$. 
				\item For each set of nodes $I_j$, consider a subset of the original network that only contains nodes in $I_j$ and edges between these nodes. Check if there is an equitable partition $I_j^{1}, ... ,I_{j}^{m_j}$, s.t. $|I_{j}^1|= ... =|I_{j}^{m_j}|$, s.t. its quotient network is $Z_{m_j}$-symmetric. If so, $I_{j}^{l}$ become candidates for fully synchronized clusters $C_p^q$. 
				\item Check that the numbers of edges coming from every fully synchronized sub-cluster of $I_j$, $I_j^{1}, ... ,I_{j}^{m_j}$ \textbf{into} every node $i$ that is not part of that cluster,  $i\notin I_j$, are equal. 
				\begin{itemize}\setlength\itemsep{0em}
					\item If the condition above is \textbf{not} satisfied, go back to step step 2 and try a new partition. In case that does not work, go to step 1 and try a new equitable partition.
					\item If the condition is satisfied, then a decoupled state is obtained. The partition $I_1,...,I_k$ assigns the nodes into splay clusters $C_p$, and its refinement $I_j^{1}, ... ,I_{j}^{m_j}$ provides the assignment into fully synchronized clusters $C_p^q$.
				\end{itemize}
			\end{enumerate}
		\end{algorithm}
	
	    Below, we provide an example of how the admissible networks for adjacency coupling may differ from those for Laplacian coupling, even for the same resulting decoupled state.
	    
	    \begin{example}\label{differences}
	    We provide examples of two 16-node networks with different coupling topologies that admit a decoupled state. In both cases, the decoupled state consists of two splay clusters. The first cluster has 3 fully synchronized sub-clusters with 4 nodes each. The second consists of 2 fully synchronized sub-clusters with 2 nodes each. 
	    
	    \cref{fig: laplacian_algo} provides an example of a network topology that admits a decoupled state for Laplacian coupling, but not adjacency coupling. The network shown in
	    \cref{fig: adjacency_algo} is an example of a network topology that admits a decoupled state for adjacency coupling, but not Laplacian coupling.
	    
	    The configuration shown in \cref{fig: laplacian_algo}(a-c) is similar to remote synchronization \cite{nicosia2013remote}: one of the fully synchronized pairs of nodes, namely, two red nodes on the bottom right of \cref{fig: laplacian_algo} have no edges directly connecting them to each other.
	    
	    \begin{figure}
	    	\includegraphics[scale=.48]{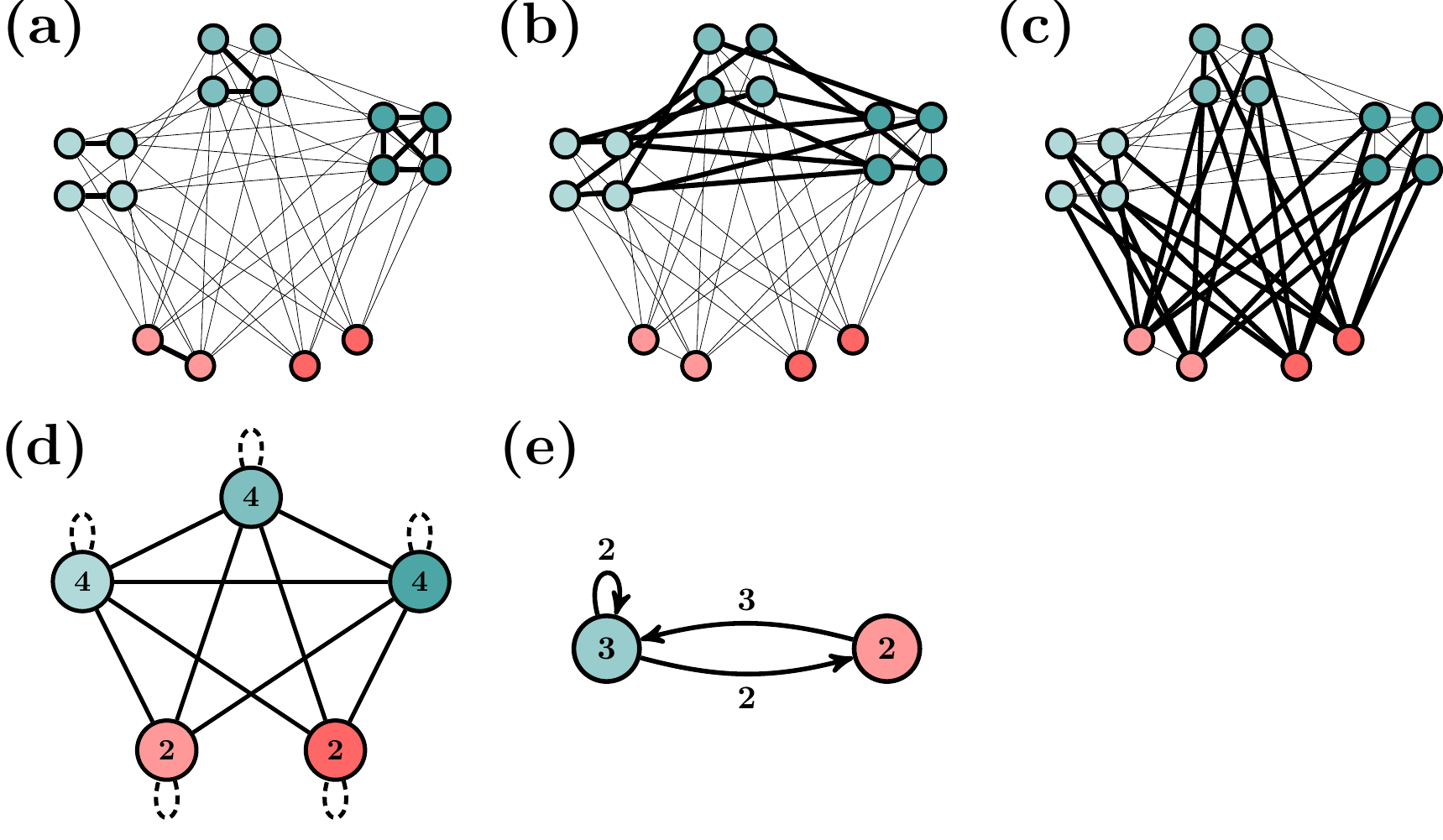}
	    	\caption{An example of a topology that allows decoupling in presence of Laplacian coupling, but not adjacency coupling.
	    		The  subfigures (a-c) have different edges of the network highlighted in bold.
	    		(a): within fully synchronized clusters, any coupling is admissible for Laplacian coupling. (b): between fully synchronized sub-clusters that are in the same splay clusters. The quotient networks are  $Z_3\times Z_2$-symmetric. (c): between splay clusters. Each red node is connected to 2 nodes in each teal sub-cluster, each teal cluster connected to 2 nodes in each red sub-cluster. (d): external equitable partition. (e): orbital partition with respect to $Z_3\times Z_2$.  The dashed lines are edges that are not mandatory.}
	    	\label{fig: laplacian_algo}
	    \end{figure}
	    \begin{figure}
	    	\includegraphics[scale=.48]{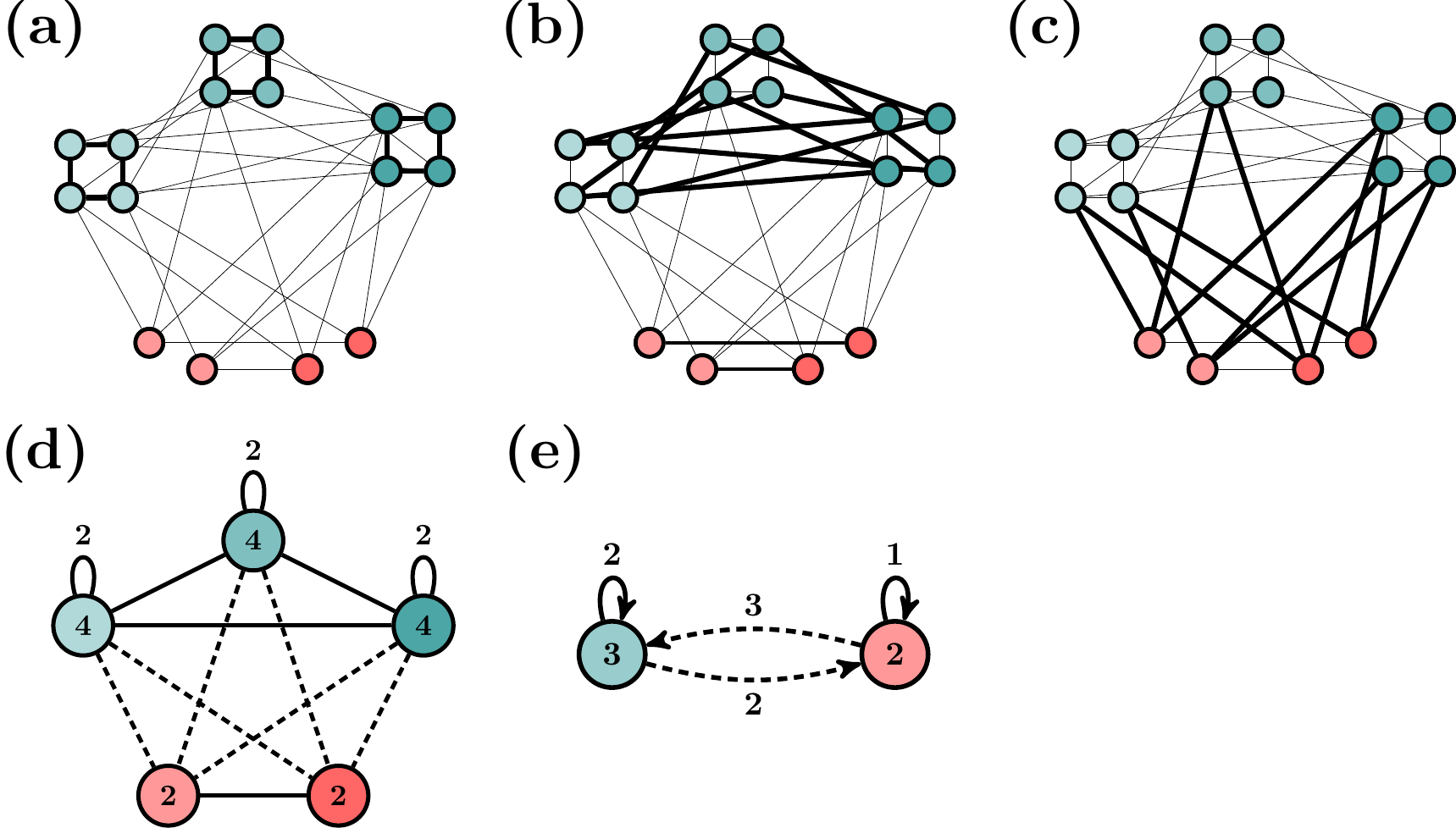}
	    	\caption{An example of a topology that allows decoupling in presence of adjacency coupling, but not Laplacian coupling.
	    		The  subfigures have different edges of the network highlighted in bold.
	    		(a): within fully synchronized clusters. (b): between fully synchronized sub-clusters, within the same splay cluster. The quotient networks are  $Z_3 \times Z_2$-symmetric. (c): between splay clusters. Each red node is connected to the same number of nodes in each teal sub-cluster, and vice versa. (d): external equitable partition. (e): orbital partition with respect to $Z_3\times Z_2$. The dashed lines are edges that are not mandatory.}
	    	\label{fig: adjacency_algo}
	    \end{figure}
	    \end{example}
    
        \subsection{Heterogeneous networks}\label{subsec: heterogeneous}
        
	    So far, we considered decoupled states arising in the networks of oscillators with homogeneous individual oscillator parameters $\psi$ and homogeneous coupling parameters $\kappa$ and $\beta$.
        However, since the interaction dynamics between different decoupled clusters sum up to zero, and since the self-interactions do not affect the state for Laplacian coupling, decoupled states can also be observed if the homogeneity assumption is relaxed. Specifically, decoupling can be observed in modular networks with different modules corresponding to different node and edge parameters, and the decoupled state is robust to small discrepancies in parameters within each modules \cite{matheny2019exotic}. Additionally, decoupling can be present in a multilayer network where each layer corresponds to a specific functional form of node dynamics and coupling parameters.

        We denote the nodal parameters by $\psi_{i}$ and the coupling parameters by $\psi_{ij}$.
        For both coupling schemes, the relaxed conditions require the following:
	   \begin{itemize}   \setlength\itemsep{0em}
		\item Nodal parameters, denoted by $\psi$, satisfy $\psi_i=\psi_j$ if $i\in C_p$ and $j\in C_p$. 
		\item  Coupling parameters between different fully synchronized sub-clusters of each decoupled splay cluster satisfy $\psi_{ij}=\psi_{kl}$ (here, $i,k\in C_p^{q}$ and $j,l\in C_p^{r}$, $q\neq r$). 
	    \end{itemize} 

     Additionally, for Laplacian coupling we require that the inter-cluster coupling parameters satisfy $\psi_{ij}=\psi_{kl}$ if $i,k\in C_p$ and $j,l\in C_q$ when $p\neq q$. If the nodes belong to the same synchronized cluster, no restrictions on the coupling strength are needed.
     
     For adjacency coupling, we require each edge parameter $\psi_{ij}$ from a fully synchronized cluster of $C_p^q$ to a fully synchronized cluster $C_r^s$ is matched by an edge $\psi_{ij}=\psi_{kl}$ of the same strength going into each of the clusters $C_r^t$ where $s\neq t$. Additionally, we require all the intra-cluster coupling parameters within fully synchronized sub-clusters to have the same strength.
     
     The conditions above can be understood both from the eigenvector approach in \cref{subsec: eigenvectors} and the more structural approach in \cref{subsec: symmetries}. The restrictions presented here set constraints on the interactions within different blocks of the weighted coupling matrix so that the heterogeneity in parameters does not affect the relevant eigenvectors, therefore still allowing the decoupled state to be present, preserving the reasoning of \cref{subsec: eigenvectors}. Re-weighting the coupling matrix does introduce different types of edges and nodes to the networks in \cref{subsec: symmetries}, but these changes do not affect the structure of the relevant quotient networks, therefore keeping the decoupled states admissible if they were admissible in a homogeneous networks \cite{aguiar2018synchronization}.

	In addition to parameter heterogeneity, we can consider heterogeneity in nodal dynamics, resulting in a multilayer matrix with each type of dynamics corresponding to a distinct layer.
	An example of such a network could be a network of Stuart-Landau oscillators coupled to a network of nanoelectromechanical oscillators, with nodal dynamics defined in \cref{eq: sl one} and \cref{eq: nems one}.
	Very similarly to the case of node parameter heterogeneity, this setup allows the decoupled state if $f_i(|z_j|,\psi)$ and $f_j(|z_j|,\psi)$ have the same functional form and parameters for $i\in C_p$ and $j\in C_p$. 
		
		\subsection{Decoupled states on lattices}\label{section: lattices}
		
		Generally, coupled periodic and chaotic oscillators placed on lattice topologies lead to a rich variety of spatio-temporal patterns, as shown in various analytic and numerical studies \cite{belykh2008cluster, lee2018twisted, golubitsky2016rigid, antoneli2007symmetry, stewart2019symmetries}. For instance, all possible balanced two-colorings for square lattices are presented in Ref. \cite{wang2004two}. 
		Since the lattice topology is highly symmetric, a variety of decoupled states is admissible on lattice networks.
		 
		The simplest example of a lattice system is an infinite 1D chain of oscillators. A state where even and odd nodes are decoupled and next nearest neighbors are in antiphase \cite{alexander1988global} is the only admissible decoupled state for such a system. This state is similar to the pattern observed in rings of $4N$ oscillators \cite{alexander1986global,matheny2019exotic,emenheiser2019decoupled}, which is a chain with periodic boundary conditions.  
		
		More complicated decoupling patterns can arise on 2D lattice coupling topologies. 
		To illustrate them, we consider a periodic $8\times 8$ square lattice, where each node has four immediate neighbors, and a periodic $8\times 8$ hexagonal lattice, where each node has six neighbors. 
		Though the analytic expressions for eigenvectors and eigenvalues of the Laplacian of periodic square and hexagonal lattices are available \cite{pozrikidis2014introduction}, their eigenvalue spectra are highly degenerate and enumerating the decoupled states based on the results of \cref{subsec: eigenvectors} is therefore nontrivial. 
		On the other hand, the approach from \cref{subsec: algorithm} can be used, but it requires forming equitable partitions (also referred to as balanced $k$-colorings) of the lattice where $k\geq 4$. 
		
\begin{figure}
			\includegraphics[scale=.44]{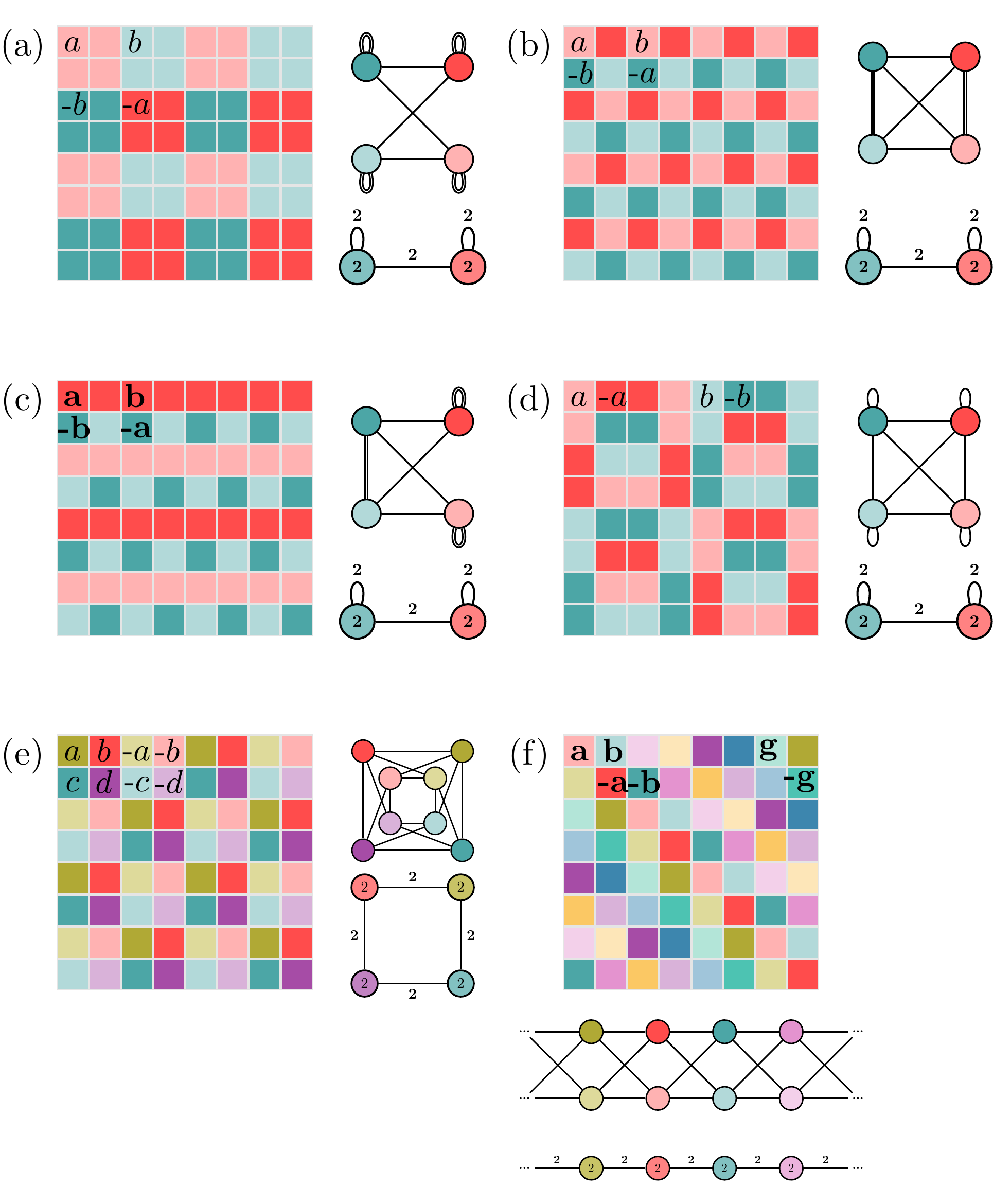}
			\caption{Decoupling in a periodic square lattice with nearest neighbor coupling (each node has 4 neighbors; in the visualization, the nodes are coupled if the edges of the squares touch). Subfigures (a-d) left: state of each node. Top right: external equitable partition defining the fully synchronized state (edges are unidirectional have weigh 1 for single line, 2 for double line). Bottom right: orbital partition defining splay clusters. Colors correspond to different splay clusters, and different brightness within each color corresponds to different fully synchronized sub-clusters. (a-d) has two splay clusters, (e) has four, (f) has $n$ for a $n\times n$ periodic lattice.}
			\label{fig: sq lattice}
		\end{figure}

\begin{figure}
			\includegraphics[scale=.44]{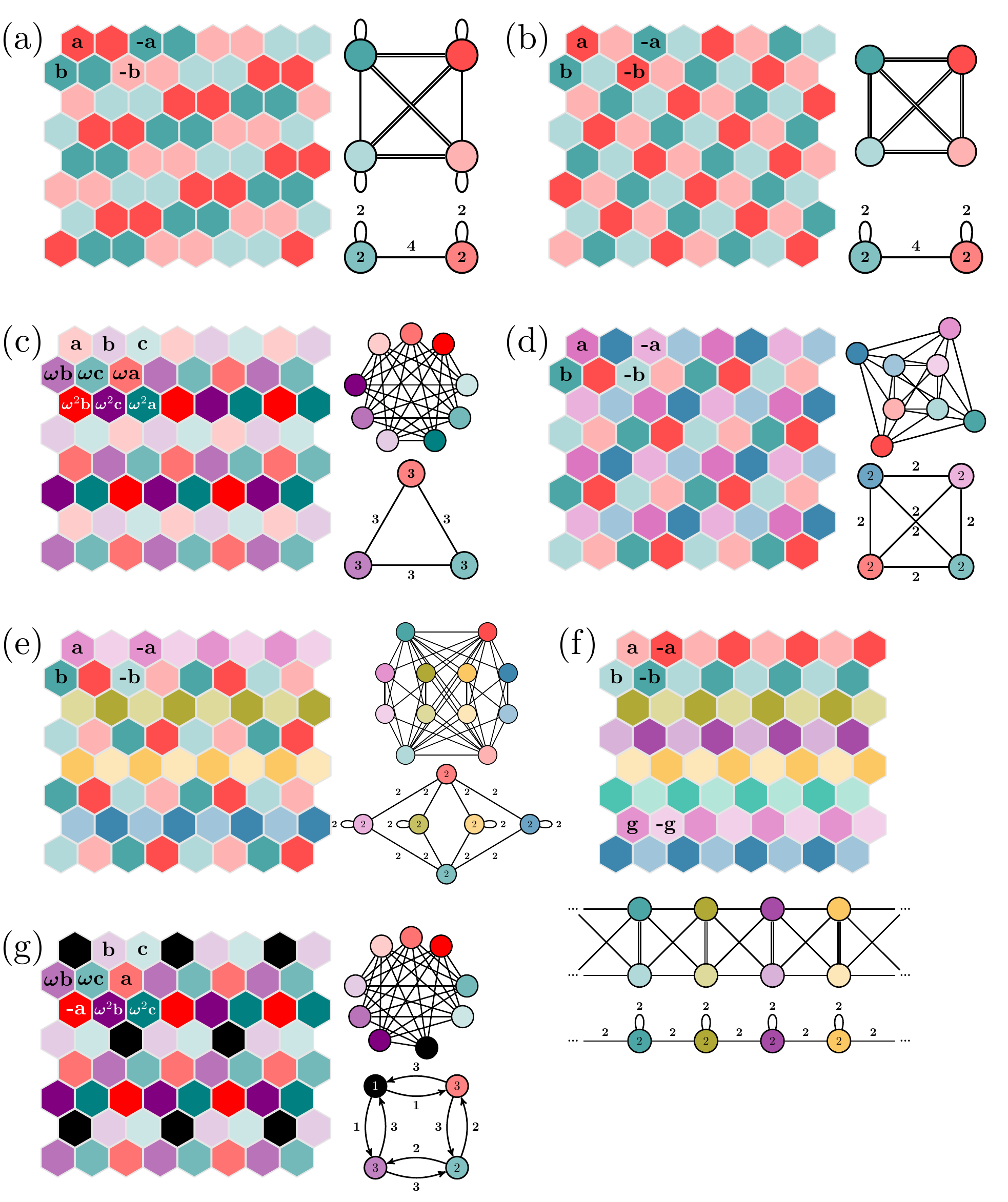}
			\caption{Decoupling in a periodic hexagonal lattice with nearest neighbor coupling (each node has 4 neighbors; in the visualization, the nodes are coupled if the edges of the hexagons touch). The meaning of parts of subfigures described under \cref{fig: sq lattice}. Subfigures (a-f) left: state of each node. Top right: external equitable partition defining the fully synchronized state (edges are unidirectional have weigh 1 for single line, 2 for double line). Bottom right: orbital partition defining splay clusters. (a-b) have two splay clusters, (c) has three, (d) four, (e) six, (f) has $n$ for a $n\times n$ periodic lattice. Subfigure (g): decoupling pattern including amplitude death.}
			\label{fig: hex lattice}
		\end{figure}

		Some of the resulting decoupling patterns, along with the quotient and orbital networks responsible for generating them, are demonstrated on \cref{fig: sq lattice,fig: hex lattice}. To keep the examples minimal, we do not introduce node or edge heterogeneity. However, the states we present can also occur if the node and edge parameters satisfy the conditions of \cref{subsec: heterogeneous}.
		
		First, we consider the square lattice case (\cref{fig: sq lattice}). We observe various patterns of decoupling, in many of which the pattern can be related to balanced $2$-colorings \cite{wang2004two}. Distinct colors represent distinct splay clusters, and different degrees of transparency represent different synchronized sub-clusters within each splay cluster. There are several partitions that lead to two decoupled clusters (\cref{fig: sq lattice} (a-d)), one that leads to four decoupled clusters (\cref{fig: sq lattice} (e)), and one that leads to eight (or, more generally, to $n$ for $n\times n$ lattices), with the phase differences between nodes on different diagonals arbitrarily defined (\cref{fig: sq lattice} (f)). In cases (d,e,f), the nodes are additionally effectively decoupled from their own cluster.  In case of adjacency coupling, this means that the oscillator frequencies and amplitudes in the decoupled state are the same as they would be in absence of any coupling.
		
		Next, we consider the hexagonal lattice coupling structure (\cref{fig: hex lattice}). We observe two distinct partitions corresponding to two decoupled clusters (\cref{fig: hex lattice} (a-b)), and one corresponding to three, four, six, and eight decoupled clusters (\cref{fig: hex lattice} (c-f)). The nodes are effectively decoupled from their own cluster in partitions (a,c-g). The state in (e) is consists of four (or, more generally, $n$ for $2n\times 2n$ lattices) separate clusters corresponding to every other row in the lattice and two decoupled clusters populating other rows. The state in (f) is similar to that in \cref{fig: sq lattice} (e), where each pair of rows is separated by an arbitrary phase difference.

		In networks of Stuart-Landau oscillators, amplitude death can coexist with the phenomena described above. 
		For instance, if any of the nonzero amplitude splay clusters is replaced with a set of dead nodes, the state is still admissible. Additionally, we present an example illustrating a different possibility on \cref{fig: hex lattice} (g). The state consists of three splay clusters and a cluster of dead nodes. Two of the splay clusters are the same as those shown on \cref{fig: hex lattice} (c). The nodes in the third cluster of \cref{fig: hex lattice} (c), shown in red, form a decoupled cluster and a cluster of dead nodes on \cref{fig: hex lattice} (g). Two decoupled clusters are characterized by winding numbers $m_{1,2}=3$, and the third cluster has $m_3=2$.
		
		In some of the patterns, all the nodes are decoupled from their own splay cluster in addition to being decoupled from all the other splay clusters. That implies the state is not linearly stable if the nodal dynamics is described by phase-only oscillator dynamics such as those of nearest neighbor Kuramoto and Kuramoto-Sakaguchi model \cite{emenheiser2019decoupled}.
		
		The analysis can be extended to higher dimensional lattices, in which even more complicated decoupling patterns could be observed. Stability of these patterns could be a subject of further investigation.

		\section{Stability calculations} \label{sec: stability}
        \subsection{General Jacobian structure}

		It is important to perform linear stability analysis of the decoupled state because it allows predicting which parameter regions will correspond to observing that state in simulations and experiments. 
		Here, we present the general outline of linear stability calculation for decoupled states in equations of the form of \cref{eq: general form} for $N=\sum\limits_{i=1}^k n_i m_i$. 
		
		The details of the stability calculation depend on the form of nodal dynamics, the decoupled state (periodic or quasiperiodic, equal or unequal winding numbers $m_i$), the type of coupling (Laplacian or adjacency matrix), and the coupling topology within splay clusters. 
		In some cases, the stability calculations can be simplified using symmetry considerations, both by block diagonalizing using subgroups of the automorphism group leading to cluster synchronization \cite{pecora2014cluster} and by block diagonalizing by additionally using the symmetries of the splay states \cite{golubitsky2003symmetry, emenheiser2019decoupled}. 
		In other cases, symmetry methods may not be applicable.
		
		Since the dynamics of each node correspond to two degrees of freedom, the Jacobian evaluated at the decoupled state is a $2N\times 2N$ matrix.
		For simplicity, we order the nodes according to the splay clusters to which they belong. 
		The full Jacobian matrix can be written as:
		\begin{align}
		J=\begin{pmatrix}
		J_{C_1,C_1}&\dots &J_{C_1,C_k}\\
		\vdots&\ddots&\vdots\\
		J_{C_k,C_1}&\dots&J_{C_k,C_k}
		\end{pmatrix},
		\end{align}
		where each block $J_{C_i,C_j}$ of size $2m_i n_i\times 2m_jn_j$ corresponds to the interactions between the splay clusters. 
		Furthermore, each of these blocks is of the form:
		\begin{align}
		J_{C_{p_1},C_{p_2}}=\begin{pmatrix}
		J_{C^{q_1}_{p_1},C^{q_1}_{p_2}}&\dots &J_{C^{q_1}_{p_1},C^{q_m}_{p_2}}\\
		\vdots&\ddots&\vdots\\
		J_{C^{q_m}_{p_1},C^{q_1}_{p_2}}&\dots &J_{C^{q_m}_{p_1},C^{q_m}_{p_2}}
		\end{pmatrix},
		\end{align}
		where the finer blocks $J_{C^{q_i}_{p_j},C^{q_k}_{p_l}}$ correspond to the interactions between the fully synchronized sub-clusters. 
		The fully synchronized sub-cluster blocks are of the form:
		\begin{align}
		J_{C_{p_1}^{q_1},C_{p_2}^{q_2}}=\begin{pmatrix}
		J_{i_1,j_1}&\dots &J_{i_1,j_{n_2}}\\
		\vdots&\ddots&\vdots\\
		J_{i_{n_1},j_1}&\dots &J_{i_{n_1},j_{n_2}}
		\end{pmatrix}.
		\end{align}
		\textcolor{black}{Here, $C_{p_1}^{q_1}=\{i_1,...,i_{n_1}\}$, and $C_{p_2}^{q_2}=\{j_1,...,j_{n_2}\}$.} 
		Finally, the Jacobian associated with each pair of oscillators is:
		\begin{align}
		J_{ij}=\begin{pmatrix}
		J_{r_i,r_j}&J_{r_i, \theta_j}\\
		J_{\theta_i,r_j}&J_{\theta_i, \theta_j}
		\end{pmatrix}=
		\begin{pmatrix}
		\dfrac{\dot{\delta r_i}}{\delta r_j}&\dfrac{\dot{\delta r_i}}{\delta \theta_j}\\
		\dfrac{\dot{\delta \theta_i}}{\delta r_j}&\dfrac{\dot{\delta \theta_i}}{\delta \theta_j}\\
		\end{pmatrix},
		\end{align}
		where $\delta r_i$ and $\delta \theta_i$ are small perturbations of the amplitude and phase of the $i$\ts{th} oscillator around the decoupled state.
		
		For instance, if $m_1=...=m_k$, the full Jacobian $J$ is a $2N\times 2N$ matrix, where $N=(n_1+...+n_k)m$. Then, for $i_1=\sum\limits_{j=1}^{p_1-1} mn_j+(p_1-1)q_1+r_1$ and $i_2=\sum\limits_{j=1}^{p_2-1} mn_j+(p_2-1)q_2+r_2$, $J_{i_1,i_2}$ represents the interactions between the oscillators in $C_{p_1}^{q_1}$ and $C_{p_2}^{q_2}$ blocks.
		
	Given a specified decoupled state, the Jacobian blocks $J_{ij}$ (where $i\in C_{p_1}^{q_1}$ and $i\in C_{p_2}^{q_2}$) on that decoupled state can be evaluated explicitly:
		\begin{align}\label{eq: stability example}
		J_{ij} &= \left\{\begin{array}{lr}
		\begin{pmatrix}
		{\partial f_{r_i}}/{\partial r_i}&\sum\limits_{i\in C_p}\kappa A_{ij} r\sin\beta\\
		\sum\limits_{i\in C_p} \kappa A_{ij}\dfrac{1}{r}\sin\beta&\sum\limits_{i\in C_p}- \kappa A_{ij} \cos\beta\\
		\end{pmatrix} 
		,\\ \text{for  } i=j;\\
		A_{ij}\kappa		
		\begin{pmatrix}
		\cos(\beta+\Delta^{q_1q_2}_{p_1p_2})&-r\sin(\beta+\Delta^{q_1q_2}_{p_1p_2})\\
		\dfrac{1}{r}\sin(\beta+\Delta^{q_1q_2}_{p_1p_2})&\cos (\beta+\Delta^{q_1q_2}_{p_1p_2})
		\end{pmatrix}
		,\\ \text{for  } i\in C^{q_1}_{p_1}, j \in C^{q_2}_{p_2},
		\end{array}
		\right . 
		\end{align}
		where we assume homogeneous coupling parameters.
		Here, for the general quasiperiodic case in case of uniform $m$, $\Delta^{q_1q_2}_{p_1p_2}(t)=\sigma_{p_1p_2}(t)+\sigma_{q_1 q_2}$, where $\sigma_{q_1 q_2}=\dfrac{2\pi}{m}\times\left((q_1-q_2)\mod m\right)$, and $\sigma_{p_1p_2}(t)=\sigma_{p_1p_2}(0)+\dot{\sigma}_{p_1p_2} t$ is a linearly evolving phase difference between the clusters. 
		
		If the Jacobian is time-independent, linear stability analysis can be performed by obtaining the largest real part of a Jacobian eigenvalue corresponding to transverse perturbations (as opposed to the neutrally stable  perturbations of phases in the direction within the state correspond to $k$ zero eigenvalues, where $k$ is the number of splay clusters in the decoupled state). If the Jacobian is a periodic matrix (namely, only two clusters have an irrational frequency ratio), the stability analysis can be performed using Floquet theory. That is always possible when only two decoupled clusters are present. If more than two clusters are present and no special restrictions are imposed on frequencies, the Jacobian is quasiperiodic and different methods are needed to perform stability analysis.

	    \subsection{States arising directly from symmetries}\label{subsec: symmetries}
	    
	    Symmetries are extremely useful in analyzing various states of equivariant networked dynamical systems \cite{golubitsky2003symmetry,pecora2014cluster}, determining  the observability and controllability of such dynamical systems \cite{whalen2015observability,Dellnitz_2016,mesbahi2019nonlinear}, as well as studying their global behaviors via transfer operators and their numerical approximations \cite{salova2019koopman}.
	    Likewise, if the decoupled state arises from symmetries alone, these symmetries can be used to assist in stability calculations using linear representation theory as shown below. 
	    Though the case of $m_i\neq m_j$ can appear as a result of symmetries, we first focus on the case of uniform $m$.
	    The symmetries of the Jacobian in that case are a combination of the symmetries associated with the orbital partition of the network, and the symmetries leading to the splay state.
	    The splay state symmetries show up as follows.
		If $i\in C^{q_1}_{p_1}$ and $j\in C^{q_2}_{p_2}$, $k\in C^{q_3}_{p_3}$ and $l\in C^{q_4}_{p_4}$, for  $\left((q_1-q_2)\mod m\right)=\left((q_3-q_4)\mod m\right)$ it is the case that $J_{ij}=J_{kl}$ if $A_{ij}=A_{kl}$ (in other words, the coupling within each block is defined by a circulant matrix). 
		This simply follows from the general form of the dynamics we consider in \cref{eq: general form}. 
		Below we present the conditions under which these symmetries (e.g., discussed in \cite{alexander1988global}) can be used to simplify the linear stability calculations.
		
		Let $M$ be the coupling matrix, and let $\Gamma$ be the automorphism group of the matrix. The group is formed by permutation matrices $P_{\gamma}$ that act by relabeling the network nodes. Let $\Sigma_{c}\subseteq \Gamma$ be its subgroup such that its orbit partitions the nodes into decoupled splay clusters. Let $\Sigma_{s}\subset \Sigma_{c}$ be the subgroup such that its orbit partitions the nodes further into fully synchronized clusters. To obtain decoupled splay clusters of equal winding number $m$, we require $\Sigma_{c}$ (and therefore $\Sigma_{s}$) to be subgroups of $(S_{n_1}\times ... \times S_{n_k})\times Z_m$. Additionally, we require $Z_m\subseteq\Sigma_{c}$. Here, $S_{n_i}$ refers to the symmetric group of degree $n_i$.
		
		Since $\Sigma_{s}$ leads to fully synchronized clusters, it can be used to block diagonalize the Jacobian according to its isotypic components (or, equivalently, irreducible representations) \cite{ pecora2014cluster}. \textcolor{black}{The symmetries of the splay states and their effect on the structure of the Jacobian, however, can also be taken into account, and the Jacobian can be block diagonalized according to the irreducible representations of $\Sigma_c$ instead, leading to a finer structure \cite{golubitsky2003symmetry, emenheiser2019decoupled}.} This results in a new coordinate system, defined by a linear transformation $T$, such that $TJT^{-1}$ is block diagonal, and thus simplifies the calculation of its associated eigenvalues (static $J$ case, periodic state) or Floquet exponents (for quasiperiodic state with linearly evolving inter-cluster phase differences). A more detailed outline of the process is presented in \cref{sec: stability 2}.
		
		We first consider how the considerations above can be applied to a previously discussed example of a ring of oscillators.
		
		\begin{example}\label{ex: ring stability}
			We consider a ring of eight oscillators (similar to \cref{ex: nems}). 
			The automorphism group of the graph is the cyclic group $\Gamma = Z_8$. 
			The subgroup of the full automorphism group, $\Sigma_c=Z_4$, corresponds to partitioning the nodes into four clusters according to its group orbit. 
			The subgroup of $\Sigma_c$, $\Sigma_s=Z_2$, partitions each of the orbits into fully synchronized clusters. 
			Using the results above, we can block diagonalize the Jacobian matrix using the irreducible representations of the symmetry group $Z_4$, going from a $16\times 16$ matrix to one with $4$ blocks, each of the size $4\times 4$. 
			A detailed analysis is presented in Ref. \cite{emenheiser2019decoupled}, where it is also shown that the symmetries of the time-dependent Jacobian are preserved for quasiperiodic states arising from the same symmetry group.
		\end{example}
	
	    Next, we present how to perform this simplification for a ``cube'' of eight oscillators. 
	    The ring and cube networks exhibit the same decoupled state, but the stability simplification process is different.
		
		\begin{example}\label{ex: cube stability}
			We now  consider a (3D) cube consisting of eight oscillators (a general discussion of decoupling in hypercubes can be found, e.g., in \cite{alexander1988global}, but stability has not been addressed previously).
			The full symmetry group of such a network is $\Gamma=S_4\times Z_2$. 
			Its subgroup $\Sigma_c=Z_2\times Z_2$ corresponds to fully synchronized clusters (shown on \cref{fig: eight}). 
			Its subgroup $\Sigma_s=Z_2$ corresponds to decoupling between even and odd nodes with antisynchronization between nodes in each cluster as shown on \cref{fig: eight}, where only the nodes in inner and outer ring are fully synchronized among each other. 
			A detailed stability calculation for an example of adjacency coupled Stuart Landau oscillators is presented in \cref{sec: stability cube}.
			Again, the simplification results in going from a $16\times 16$ matrix to $4$, $4\times 4$ blocks, but the linear transformation $T$ is different from the one in \cref{ex: ring stability}. We focus on the case of \textbf{identical oscillators} for simplicity and consider adjacency and Laplacian coupling.
			\begin{figure} 
				\includegraphics[scale=.72]{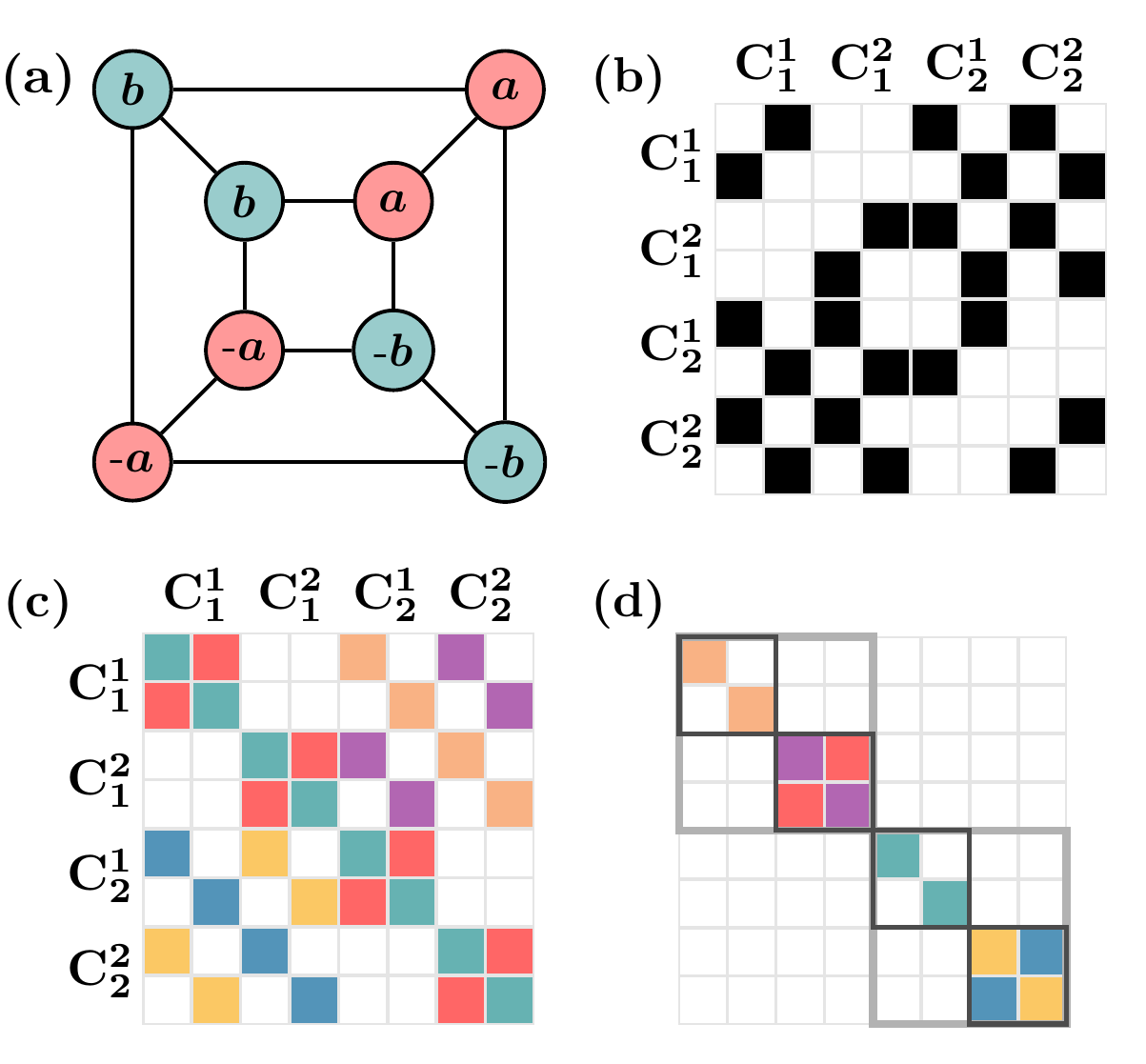}
				\caption{Example of a state on an eight-node coupled network and its Jacobian structure. (a) State of a network, where colors represent different decoupled clusters. (b) Adjacency matrix structure. (c-d) Jacobians in node and symmetry coordinates respectively. Colors represent different numerical values. Each colored block corresponds to a $2\times 2$ matrix for amplitude and phase dynamics. (d) Dark gray lines correspond to Jacobian blocks according to fully synchronized cluster symmetries. Black corresponds to the finer structure obtained by decomposing according to $\Sigma_c$.}
				\label{fig: eight}
			\end{figure}
	
		\begin{figure*}[t]{}
			\includegraphics[scale=1.3]{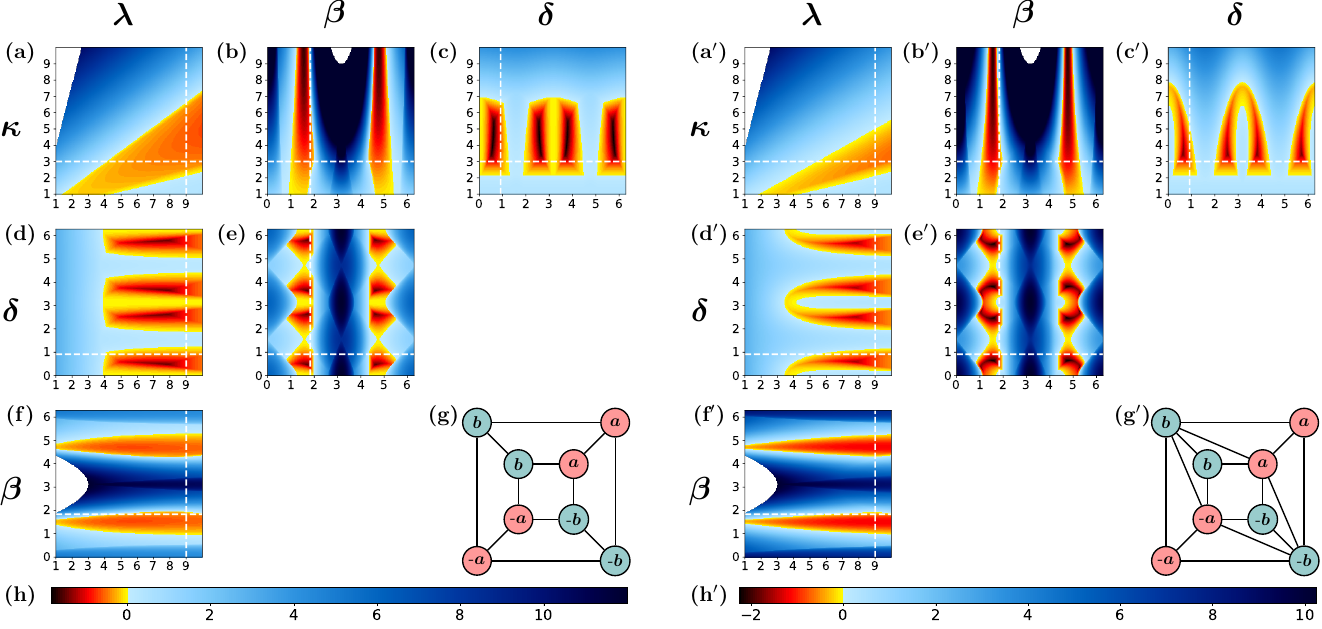}
			\caption{Linear stability of the decoupled state discussed in \cref{ex: cube stability} and \cref{sec: stability cube} for adjacency coupling. Left (a-h): coupling on a cube. Right: (a$'$-h$'$): coupling on a cube with additional edges. Subfigures $(g)$ and $(g')$ in the lower right corner of the stability plots show the coupling topologies linear stability was calculated for. Note that all the edges and all the nodes are assumed to have equal coupling and individual parameters. Largest transverse Jacobian eigenvalue is plotted for a set of parameters $\lambda$, $\sigma$, $\beta$, and $\delta$ (the angle between two splay clusters). 
			Each subplot corresponds to the case where two of these parameters are fixed, and two are varied.
			Fixed parameters are selected from $\lambda=9$, $\sigma=3$, $\beta=7\pi/12$, and $\delta=7\pi/24$.	
			Dashed white lines correspond to these fixed parameter values and can be used to guide comparing the subplots.
			Colors represent the magnitude of the real part of the maximum transverse eigenvalue of the Jacobian (blue for positive eigenvalue or linearly unstable solution, yellow and red for negative eigenvalue or linearly stable solution).
			The corresponding color bar is shown at the bottom of the figure.
			White spaces on the plots correspond to the parts of the space where there is no solution corresponding to real amplitude. The subfigures represent the following:
		    (a) and (a$'$) $\lambda$ vs $\kappa$;
		    (b) and (b$'$) $\beta$ vs $\kappa$;
		    (c) and (c$'$) $\delta$ vs $\kappa$;
		    (d) and (d$'$) $\lambda$ vs $\delta$; 
		    (e) and (e$'$) $\beta$ vs $\delta$; 
		    (f) and (f$'$) $\lambda$ vs $\beta$; 
		    (e) and (e$'$) colormap.
		     \label{fig: cube stability_A}
		}
		\end{figure*}
	
		\begin{figure*}[t]{}
		\includegraphics[scale=1.3]{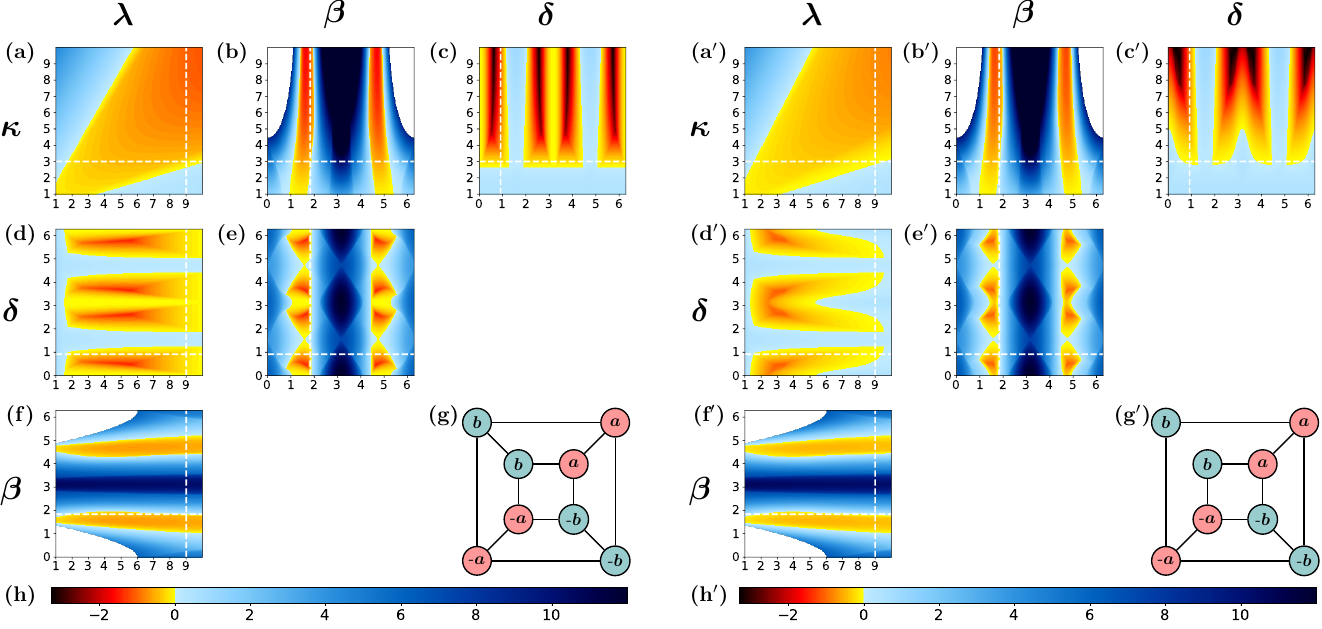}
		\caption{
		Linear stability of the decoupled state discussed in \cref{ex: cube stability} and \cref{sec: stability cube} for Laplacian coupling. Left (a-h): coupling on a cube. Right: (a$'$-h$'$): coupling on a cube with a removed edge. 
		For fixed parameter values and subplot meanings see captions of \cref{fig: cube stability_A}.
		\label{fig: cube stability_L}
		}
	    \end{figure*}
		\end{example}

        \subsection{Beyond symmetries}
        
        Isotypic component decomposition is not directly applicable if the state does not appear as a result of symmetries. However, the stability analysis can still be simplified \cite{sorrentino2016complete, siddique2018symmetry, zhang2020symmetry}.
        Here, we present a low-dimensional example where the analysis can be performed even without the extra simplifications.
        
        We modify the network in \cref{ex: cube stability} to break the original network symmetries in a way that keeps the decoupled state admissible. The original topology is shown in \cref{fig: cube stability_A} (g). 
        We demonstrate these topology modifications in \cref{fig: cube stability_A} (g$'$) and \cref{fig: cube stability_L} (g$'$).  The coupling topology of \cref{fig: cube stability_A} (g$'$) corresponds to %the coupling topology obtained by 
        adding edges between different decoupled clusters such that the inter-cluster coupling terms still add up to zero. 
        In this setting, the decoupled state can still appear in presence of adjacency coupling. We present its stability for various parameter regimes on \cref{fig: cube stability_A}(a$'$-f$'$). 
        The regions of parameter space where the state is stable are similar to those related to symmetric coupling, illustrated in \cref{fig: cube stability_A}(a-g), though the details of the stability region boundaries differ. 
        Similarly, we expect the modified topology to modify the shape of the state's basin of attraction once the parameters are fixed. 
        
        We also break the symmetry of the network in a way that makes the state admissible for Laplacian coupling by deleting an edge between two nodes in a fully synchronized cluster, as shown in \cref{fig: cube stability_L} (g'). The resulting stability diagrams are shown in \cref{fig: cube stability_L} (a$'$-f$'$). The stability diagrams show that symmetry breaking affects the shape of the stability regions, but preserves the possibility of observing these states in experiments, e.g., the experimental realization of networks of Stuart-Landau oscillators described in Ref. \cite{karakaya2019fading}.
        
        \iffalse
        Recently, it was shown that the it is possible to block diagonalize the stability matrix even if the synchronization patterns arise from external equitable partitions, and not orbital partitions. In that case, simultaneous block diagonalization can be performed using the algorithm in Ref. \cite{zhang2020symmetry}. That algorithm performs simultaneous block diagonalization of the Laplacian (or adjacency matrix) and the indicator matrices for fully synchronized clusters.
        
        \cref{nems2} demonstrates a decoupled state that is not directly a consequence of symmetries. Using the algorithm from Ref. \cite{zhang2020symmetry} allows to find the simultaneous block diagonalization of the Laplacian matrix and four indicator matrices with diagonal elements corresponding to fully synchronized nodes.
        \fi
		
		\iffalse
		\section{Sensitivity to parameter mismatches: synchronization and stability}\label{sec: sensitivity}
		
	    Ref. \cite{sorrentino2016approximate, golubitsky2016rigid}.
	   
	    In order to study the 
		
	    The strategy above is directly applicable to the cases when all the splay clusters arise from orbital partitions both for fully synchronized clusters and splay clusters, and all the splay clusters are described by $Z_m$ (which can be the case even in presence of intracluster parameter mismatches).

		\begin{align}
		\dot{x}_i(t)=F(x_i,m_i)+\sigma\sum A_{ij}H(x_j)
		\end{align}
		\begin{align}
		A=A^{sym}+\delta A
		\end{align}
		\begin{align}
		\delta X=A^{sym}+\delta A
		\end{align}
		 
		\fi
		
		\section{Discussion}\label{sec: discussion}
  
  Synchronization phenomena, especially understanding the origins and stability of non-trivial synchronization patterns, are of great interest for both theoretical and practical reasons. Here we focus on states of synchronization that result from the canceling of terms in the dynamical equations of evolution that lead to decoupling of often physically coupled oscillators. Such states show intriguing, emergent long-range order, such as next-nearest neighbor antiphase synchronization with seeming independence between physically coupled nearest neighbors, as shown for the ring of 8 oscillators used as an example throughout this work (e.g.,~\cref{fig: clusters1} (e)).   Here we consider the broad class of possible decoupled states that are accessible given the underlying network topology and the nature of the coupling between oscillators.  Note that specific decoupled states have been studied in the literature previously based on symmetry considerations~\cite{alexander1988global, alexander1986global,matheny2019exotic}, but a unifying treatment of the class of states and their stability properties did not previously exist.  Our work accomplishes this and also reveals that network symmetries alone are insufficient for identifying all possible decoupled states that a system can support, since the symmetries of the quotient network do not necessarily translate to the symmetries of the full network. 
  
        Specifically, we analyze the continua of decoupled states in networks of linearly coupled rotationally invariant limit cycle oscillators. 
        We show that the eigendecomposition of the coupling matrix can reveal which decoupled states are admissible. 
        We also formulate the admissibility criteria in the language of equitable and orbital partitions. This takes into account the balanced equivalence relations of the network as well as the symmetries of the associated quotient network and makes connections to symmetry groupoid and cluster synchronization literature.
        We demonstrate how various forms of decoupled states can arise in systems such as lattices of oscillators with periodic boundary conditions, and how partial amplitude death can a be part of these decoupled states.
		
		Some of the most commonly considered cases of coupling for networked dynamical systems are the adjacency and Laplacian coupling schemes.
		Knowing the precise form of the coupling interactions lets us study the effect of the network structure on decoupling admissibility in great detail. 
	    We find that the Laplacian coupling scheme admits more flexibility in coupling between fully synchronized nodes, since the time evolution is not affected by the edges within fully synchronized clusters.
		Adjacency coupling, on the other hand, poses restrictions on the edges within fully synchronized clusters. 
		However, it allows more flexibility in connections between different splay clusters, since the only condition that has to be imposed for the state to be admissible is that the contributions to each node from the splay clusters it does not belong to cancel out.
		
		If the decoupled state is a direct result of network symmetries (as opposed to the symmetries of the quotient network), it is admissible for both adjacency and Laplacian coupling schemes.
		Additionally, in that case the stability analysis can be simplified by block diagonalizing the Jacobian according to the irreducible representations of the symmetry group related to the state.
		We show how finer block sizes can be achieved by considering the symmetries beyond the automorphism group of the coupling matrix. 
		We pick a simple eight node cube network, as well as those obtained from it by symmetry breaking edge addition and deletion, to illustrate the numerical linear stability analysis and determine what parameter regions can allow the observation of the decoupled state in simulations and experiment.
		We show that the stability regions are relatively robust to edge perturbations.
		
        Our analysis of decoupled states is generalizable. For instance, it can be easily extended to networks with directed coupling, or even multilayer networks. In multilayer networks, decoupling could be present in one or more layers, or correspond to layers being decoupled from each other. In addition, the definition of the state itself can be extended to include dead nodes decoupled from all the other nodes, or, in case of Laplacian coupling, the nodes that are only attached to their cluster and therefore do not change the state of all the other nodes.
        In future work, it would be interesting to investigate the stability of decoupling combined with amplitude death, as that coexistence may be allowed for a larger set of network topologies. 
		Moreover, decoupling is robust with respect to small parameter mismatch between individual oscillator parameters, as shown in experiments \cite{matheny2019exotic}.
		That robustness, as well as the fact that the state can be observed for diverse coupling topologies, e.g., modular coupling with all-to-all coupling between the modules, could mean decoupling can occur in natural systems such as biological networks and be related to behaviors such as remote mediated synchronization in the brain \cite{qin2020mediated}.
		Since the stability analysis in presence of parameter heterogeneity exhibits sensitive dependence on the parameter values corresponding to islands of stability \cite{emenheiser2019decoupled}, such analysis is an important step towards understanding and predicting decoupling in experimental and natural systems.

		\section{Acknowledgments}
		
	The authors thank Jeffrey Emenheiser for helpful conversations.	We gratefully acknowledge support from the US Army Research Office MURI Award No. W911NF-13-1-0340.

		\appendix
		\newpage 
		
		\section{Decoupling in Stuart Landau oscillators}\label{app: SL}
		
		Here, we provide the form of the decoupled solution for Stuart-Landau oscillator networks and show how it leads to phenomena that have not been explicitly discussed in Stuart-Landau literature. 
		The dynamics in presence of adjacency coupling can be expressed as:
		\begin{align}\label{eq: SL}
		\dot{z}_j=(\lambda_j+i\omega_j -|z_j|^2)z_j+\sum\limits_k M_{jk} \kappa_{jk} e^{i\beta_{jk}}z_k.
		\end{align}
		Here, $M$ can be an adjacency matrix or a Laplacian matrix.
		Equivalently, in phase-amplitude coordinates:
		\begin{align}\label{eq: SL and}
		\begin{gathered}
		\dot{r}_j=(\lambda_j-r_j^2)r_j+ \sum\limits_{k}M_{jk}\kappa_{jk} r_k\cos(\beta_{jk} + \theta_k-\theta_j),\\
		\dot{\theta}_j=\omega_j + \sum_{k} M_{jk} \kappa_{jk} \dfrac{r_k}{r_j}\sin(\beta_{jk} + \theta_k-\theta_j).
		\end{gathered}
		\end{align}
		
		If $j\in C_p$, the only nodes that have effect on $j$ also belong to $C_p$. 
		Let $\tilde{\lambda}=\sum\limits_{k\in C_p}A_{jk}\kappa_p\cos\left(\beta_{p} + \theta_k-\theta_j\right)$ and $\tilde{\omega}=\sum\limits_{k\in C_p}A_{jk}\kappa_p\sin\left(\beta_{p} + \theta_k-\theta_j\right)$.
		The adjacency coupling dynamics on the decoupled state can be expressed as:
        \begin{align}
        r_j &= \sqrt{\lambda_j+\tilde{\lambda}_p},\\
        \theta_j&=\theta_j(0)+(\omega_j +\tilde{\omega})t.
        \end{align}
        Here, the parameters $\beta_p$ and $\kappa_p$ are the in-cluster coupling parameters, $\theta_j(0)$ satisfy the decoupled state conditions.
        
        For instance, if the oscillators are not directly coupled to the any other ones in their group $C_p$ (e.g., ring topology in \cref{nems1}), and the decoupled state is admissible, the solution takes form:
        \begin{align}\label{eq: natural freq}
        r_j&=\sqrt{\lambda_j},\\
        \dot{\theta}_j&=\omega_j.
        \end{align}
        The oscillators move at their natural frequencies in the same manner they would in absence of coupling. Their amplitudes and frequencies are uniform within splay clusters, but may differ between the clusters if parameter heterogeneity is present.
        
        For Laplacian coupling with $M=A-D$, the dynamics is:
        \begin{align}
        r_{j\in C_p} &= \sqrt{\lambda_p+\tilde{\lambda}_p-\sum\limits_{C_p}A_{jk}\kappa_p\cos\beta},\\
        \theta_{j\in C_p}&=\theta_j(0)+\left(\omega_p +\tilde{\omega}-\sum\limits_{C_p}A_{jk}\kappa_p\sin\beta\right)t.
        \end{align}
        Here, each coupling edge introduces a shift in oscillator amplitudes and phases.
        
        \iffalse
        \as{Not sure if we need this here.}\textcolor{red}{A zero amplitude node would not change an eigensolutions as long as it is \textit{decoupled} from the nodes it is attached to \cite{poel2015partial}. In our case, this could mean the nodes in an entire decoupled cluster.}
        \fi
		
		\section{Adding state symmetries to cluster synchronization} \label{sec: stability 2}
		
		As shown in the \cref{subsec: symmetries}, the Jacobian of the dynamics on decoupled states commutes with the permutations generated by the actions of the group $\Sigma_c$, the orbits of which form \textit {splay} clusters (not just the subgroup $\Sigma_s$ and associated fully synchronized sub-clusters). 
		
		To block diagonalize the Jacobian matrix, we take the following steps. First, we find the projection operators $T^{(l)}$ from the following expression:
		\begin{align}
		T^{(l)}=\dfrac{d^{(l)}}{h}\sum\limits_{\kappa}\chi_{\kappa}^{(l)}\sum\limits_{g\in\kappa}R_g
		\end{align} 
		Here, $d^{(l)}$ is the dimension of the $l$\ts{th} irreducible representation of $\Sigma_c$, $h$ is the size of the symmetry group $\Sigma_c$, $\kappa$ is a conjugacy class, $\chi_{\kappa}$ are the characters corresponding to a conjugacy class $\kappa$ and an irreducible representation $l$, and $R_g$ are the linear representations of group elements $g\in\Sigma_c$ in a form of permutation matrices.
		
		Stacking the eigenvectors ot $T^{(l)}$ provides a projection matrix $T$, which can be used to transform the Jacobian into a block diagonal form: $J_{BD}=TJT^{-1}$. 
		A more detailed description of the process as applied to clusters of identically synchronized oscillators can be found in recent cluster synchronization literature \cite{pecora2014cluster}, and the process of obtaining the projection can be simplified using computational group theory tools developed to address this problem \cite{hagerstrom2014network}. 
		
		\section{Detailed example of stability calculations} \label{sec: stability cube}
		Here, we provide explicit stability matrix block diagonalization for \cref{ex: ring stability} and a periodic solution. We consider the Stuart Landau oscillator dynamics:
        \begin{align}
		\dot{z}_j=(\lambda+i\omega -|z_j|^2)z_j+\sum\limits_kM_{jk}\kappa e^{i\beta}z_k.
		\end{align}
		
		The adjacency matrix $A$ and the form of the decoupled state of interest $z_{\text{dc}}$ are:
		
		\begin{align} \label{eq: adjacency 8}
		\renewcommand*{\arraystretch}{.4}
		A = \left(
		\begin{array}{cc|cc|cc|cc}
		0&1&0&0&1&0&1&0\\
		1&0&0&0&0&1&0&1\\
		\hline
		0&0&0&1&1&0&1&0\\
		0&0&1&0&0&1&0&1\\
		\hline
		1&0&1&0&0&1&0&0\\
		0&1&0&1&1&0&0&0\\
		\hline
		1&0&1&0&0&0&0&1\\
		0&1&0&1&0&0&1&0
		\end{array}
		\right),~~
		z_{\text{dc}}=
		\begin{pmatrix}
		a\\
		a\\
		- a\\
		- a\\
		b\\
		b\\
		- b\\
		- b
		\end{pmatrix}.
		\end{align}
		where $|a|=|b|$. In case of adjacency coupling, $M=A$, and $M=A-D$ for Laplacian coupling. The time evolution in case of adjacency coupling is defined by:
		\begin{align}\label{eq: r_a}
		\begin{gathered}
		r_j(t) = \sqrt{\lambda+\kappa\cos\beta},\\
		\dot\theta_j(t)=\omega + \kappa\sin\beta.
		\end{gathered}
		\end{align}
		If the coupling is Laplacian,
		\begin{align}\label{eq: r_l}
		\begin{gathered}
		r_j(t) = \sqrt{\lambda-2\kappa\cos\beta},\\
		\dot\theta_j(t)=\omega -2 \kappa\sin\beta.
		\end{gathered}
		\end{align}
		
		Let $\Sigma_c=Z_2\times Z_2$ be the group defining the splay clusters, as discussed in \cref{sec: stability}. The group has two commuting generators (we denote them by $\alpha$ and $\beta$, and the identity element by $e$), and the corresponding representations acting on the coupling matrix are:
		\begin{align} \label{eq: P cluster}
		R_{\alpha} &= 
		I_{4\times 4}\otimes
	    \renewcommand*{\arraystretch}{.6}\begin{pmatrix}
		0&1\\
		1&0\\
		\end{pmatrix},\\ \label{eq: P splay}
		R_{\beta} &= 
		I_{2\times 2}\otimes
	    \renewcommand*{\arraystretch}{.6}\begin{pmatrix}
		0&1\\
		1&0\\
		\end{pmatrix}
		\otimes I_{2\times 2} .
		\end{align}
		All elements of this group commute with the adjacency matrix $A$ (as well as $L$). The group has four irreducible representations, with characters presented in the table below.
		
		\begin{center}
			\begin{tabular}{ |c|c|c|c|c| } 
				\hline
				& $e$ & $\alpha$ &$\beta$&$\alpha\beta$\\ 
				\hline
				$\chi_{11}$ &1&~1&~1&~1\\ 
				$\chi_{12}$ &1&-1&~1&-1\\ 
				$\chi_{21}$ &1&~1&-1&-1\\ 
				$\chi_{22}$ &1&-1&-1&~1\\  
				\hline
			\end{tabular}
		\end{center}   
		
		The projections onto the isotypic component basis are then defined by their nontrivial eigenvectors:
		\begin{align}
		T_{11}&=
			    \renewcommand*{\arraystretch}{.6}
		\begin{pmatrix}
		1&1&1&1&0&0&0&0\\
		0&0&0&0&1&1&1&1\\
		\end{pmatrix},\\
		T_{12}&=
					    \renewcommand*{\arraystretch}{.6}
		\begin{pmatrix}
		1&1&-1&-1&0&0&0&0\\
		0&0&0&0&1&1&-1&-1\\
		\end{pmatrix},\\
		T_{21}&=
					    \renewcommand*{\arraystretch}{.6}
		\begin{pmatrix}
		1&-1&1&-1&0&0&0&0\\
		0&0&0&0&-1&1&-1&1\\
		\end{pmatrix},\\
		T_{22}&=
					    \renewcommand*{\arraystretch}{.6}
		\begin{pmatrix}
		1&-1&-1&1&0&0&0&0\\
		0&0&0&0&-1&1&1&-1\\
		\end{pmatrix}.
		\end{align}
		The transformation matrix can be obtained by vertically stacking these projection matrices. 
		
		To illustrate the stability calculation, we first provide the form of the Jacobian evaluated at the decoupled state:
		\begin{align} \label{eq: adjacency 8}
		J = \left(
		\begin{array}{cc|cc|cc|cc}
		J_D        &J^{11}_{11}&0          &0          &J^{11}_{12}&0          &J^{12}_{12}&0          \\
		J^{11}_{11}&J_D          &0          &0          &0          &J^{11}_{12}&0          &J^{12}_{12}\\
		\hline
		0          &0          &J_D        &J^{11}_{11}&J^{12}_{12}&0          &J^{11}_{12}&0          \\
		0          &0          &J^{11}_{11}&J_D        &0          &J^{12}_{12}&0          &J^{11}_{12}\\
		\hline
		J^{11}_{21}&0          &J^{21}_{21}&0          &J_D        &J^{11}_{11}&0          &0          \\
		0          &J^{11}_{21}&0          &J^{21}_{21}&J^{11}_{11}&J_D        &0          &0          \\
		\hline
		J^{21}_{21}&0          &J^{11}_{21}&0          &0          &0          &J_D        &J^{11}_{11}\\
		0          &J^{21}_{21}&0          &J^{11}_{21}&0          &0          &J^{11}_{11}&J_D
		\end{array}
		\right).
		\end{align}
		
        Here, each element is a $2\times 2$ block. Except for the self-interaction blocks, $J_D$ (denoted by $J_D^A$ and $J_D^L$ respectively), the blocks are the same for adjacency and Laplacian coupling. The blocks are defined by:
		\begin{align}\label{eq: jacobians}
		\begin{gathered}
		J_D^A=
		\begin{pmatrix}
		\lambda-3r^2&\kappa r\sin\beta\\
		-\kappa/r\sin\beta&-\kappa \cos\beta\\
		\end{pmatrix},\\
		J_D^L=
		\begin{pmatrix}
		\lambda-3r^2-3\kappa \cos\beta&\kappa r\sin\beta\\
		-\kappa/r\sin\beta&-\kappa \cos\beta\\
		\end{pmatrix},\\
		J_{11}^{11}=\kappa
		\begin{pmatrix}
		\cos\beta&- r\sin\beta\\
		r \sin \beta& \cos\beta\\
		\end{pmatrix},\\
		J_{12}^{11}=\kappa
		\begin{pmatrix}
		\sin\delta& r\cos\delta\\
		r\cos\delta&-\sin\delta\\
		\end{pmatrix},~~
		~~J_{12}^{12}=-J_{12}^{11},\\
		J_{21}^{11}=\kappa
		\begin{pmatrix}
		- \sin\delta& r\cos\delta\\
		r\cos\delta&\sin\delta\\
		\end{pmatrix},~~
		~J_{21}^{21}=-J_{21}^{11},
		\end{gathered} 
		\end{align}
		where $\delta=\delta_{12}+\beta$, and $J_D^A$ and $J_D^L$ refer to $J_D$ in case of adjacency and Laplacian coupling respectively, and the values of $r$ can be obtained from \cref{eq: r_a} and \cref{eq: r_l} for the adjacency and Laplacian cases respectively.
		Then $J_{BD} = J_1\oplus J_2\oplus J_3\oplus J_4$, where:
		\iffalse \begin{align}
		\begin{gathered}
		J_1 = \begin{pmatrix}
		J_D+J_{11}&J^{11}_{12}+J^{12}_{12}\\
		J^{11}_{21}+J^{21}_{21}&J_D+J_{11}\\
		\end{pmatrix}\\
		J_2 = \begin{pmatrix}
		J_D+J_{11}&J^{11}_{12}-J^{12}_{12}\\
		J^{11}_{21}-J^{21}_{21}&J_D+J_{11}\\
		\end{pmatrix}\\
		J_3 = \begin{pmatrix}
		J_D-J_{11}&-J^{11}_{12}-J^{12}_{12}\\
		-J^{11}_{21}-J^{21}_{21}&J_D-J_{11}\\
		\end{pmatrix}\\
		J_4 = \begin{pmatrix}
		J_D-J_{11}&-J^{11}_{12}+J^{12}_{12}\\
		-J^{11}_{21}+J^{21}_{21}&J_D-J_{11}\\
		\end{pmatrix}.
		\end{gathered}
		\end{align}
		\fi
		\begin{align}
		\begin{gathered}
		J_1 = \begin{pmatrix}
		J_D+J_{11}&0\\
		0&J_D+J_{11}\\
		\end{pmatrix}\\
		J_2 = \begin{pmatrix}
		J_D+J_{11}&J^{11}_{12}-J^{12}_{12}\\
		J^{11}_{21}-J^{21}_{21}&J_D+J_{11}\\
		\end{pmatrix}\\
		J_3 = \begin{pmatrix}
		J_D-J_{11}&0\\
		0&J_D-J_{11}\\
		\end{pmatrix}\\
		J_4 = \begin{pmatrix}
		J_D-J_{11}&-J^{11}_{12}+J^{12}_{12}\\
		-J^{11}_{21}+J^{21}_{21}&J_D-J_{11}\\
		\end{pmatrix}.
		\end{gathered}
		\end{align}
		Additional zero structure within $J_1$ and $J_3$ arises from the form of \cref{eq: jacobians}. The eigenvalues of $J_1$ and $J_3$ can be computed analytically. This reduced the size of the Jacobian blocks and speeds up the stability computations.
		
		We present a linear stability diagram for such a system as a function of parameter values $\lambda$, $\kappa$, $\beta$, and $\delta_{1,2}$ in \cref{fig: cube stability_A} and \cref{fig: cube stability_L} for adjacency and Laplacian cases respectively. Colors show the value of the maximum transverse Lyapunov exponent $\eta_{\max}$. We note that the symmetries of equations lead to $\text{Re}(\eta(\delta))_{\max}=\text{Re}(\eta(\pi/2-\delta))_{\max}$,  $\text{Re}(\eta(\delta))_{\max}=\text{Re}(\eta(-\delta))_{\max}$, and $\text{Re}(\eta(\beta))_{\max}=\text{Re}(\eta(-\beta))_{\max}$, where $\eta$s stand for the Jacobian eigenvalues, and all the parameters not stated in parentheses (e.g., $\lambda,\omega,\kappa,\delta_{12}$ for $\eta(\beta)$) are kept constant.
    
    \iffalse\section{Other models}
    \begin{align}
    \dot{x}_i = f(x_i) + \sum g(x_1,...,x_N)
    \end{align}
    Let $f$ and $g$ share a period $T$. Then the decoupled state is $\{a, a(T/n), ..., a(T(n-1)/n), b, b(T/n), ..., b(T(n-1)/n)\}$.
    
    $\sum f(0)+f(T/n)+...+f(T(n-1)/n)=0$

    \section{Linear stability of phase only models} \label{app: kuramoto}
	
	In Kuramoto-Sakaguchi model, the phase of each oscillator evolves according to:
	\begin{align}
    \dot{\theta}_j=\omega_j+\sum_{k \sim j}\sin(\theta_k-\theta_j+\delta).
	\end{align}
	Here, the diagonal elements of the Jacobian are:
	\begin{align}
	J_{jj}=-\sum_{k\sim j}\cos(\theta_k-\theta_j+\delta).
	\end{align}
	Here, $\sum\limits_{k\sim j}\cos(\theta_k-\theta_j+\delta)=0$ if $j\in C_p$ and $k\notin C_p$. If there are no other nodes, the state is linearly unstable since the trace is zero. On the other hand, if there are other edges (if $j\in C_p$ and $k\in C_p$):
	\begin{align}
	\sum\limits_{k\sim j}\cos(\theta_k-\theta_j+\delta)=l_1\cos\delta+l_2\sum_{k\sim j}\cos(2\pi/m+\delta).
	\end{align}
	The state is necessarily unstable if $l_1\cos \delta + l_2\sum\limits_{k\sim j}\cos(2\pi/m+\delta)<0$.
	\fi

	\newpage
	\bibliography{biblio}
	
	\end{document}